\documentclass{article}%
\usepackage{amsmath}
\usepackage{amsfonts}
\usepackage{amssymb}
\usepackage{graphicx}
\usepackage{comment}
\usepackage{hyperref}%
\providecommand{\U}[1]{\protect\rule{.1in}{.1in}}

\newtheorem{theorem}{Theorem}

\newtheorem{remark}[theorem]{Remark}

\begin{document}

\title{Libor model with expiry-wise stochastic volatility and
displacement\thanks{Weierstrass Institute for Applied Analysis and
Stochastics, Mohrenstr. 39, 10117 Berlin, Germany.
\texttt{{Marcel.Ladkau/John.Schoenmakers/Jianing.Zhang@wias-berlin.de}}.
Supported by the DFG Research Center \textsc{Matheon} `Mathematics for Key
Technologies' in Berlin. 
\qquad  The authors are grateful to Dr. Suso Kraut and Dr.
Marcus Steinkamp (HSH Nordbank) for many stimulating discussions. }}
\author{Marcel Ladkau, John Schoenmakers, Jianing Zhang}
\maketitle

\begin{abstract}
We develop a multi-factor stochastic volatility Libor model with
displacement, where each individual forward Libor is driven by its own
square-root stochastic volatility process. The main advantage of this approach
is that, maturity-wise, each square-root process can be calibrated to the
corresponding cap(let)vola-strike panel at the market. However, since
even after freezing the Libors in the drift of this model, the Libor dynamics
are not affine, new affine approximations have to be developed in order to
obtain Fourier based (approximate) pricing procedures for caps and swaptions.
As a result, we end up with a Libor modeling package that allows for efficient
calibration to a complete system of cap/swaption market quotes that performs
well even in crises times, where structural breaks in vola-strike-maturity
panels are typically observed.

\end{abstract}



\noindent\emph{JEL Classification Code}: G1

\section{Introduction and summary}

The framework of Libor interest rate modeling, initially developed by
\cite{MSS}, \cite{BGM}, and \cite{J2} almost two decades ago, is still
considered to be the universal tool for evaluation of structured interest rate
products. One of the main reasons for this is the great flexibility of the
Libor framework: It allows to include many sources of randomness of different
type, such as Brownian motions, L\'evy processes, or even more general
semimartingales (see e.g. \cite{J1}). Subsequently, these random sources may
be connected with different types of volatility structures, such as stochastic
volatility, local volatility, or deterministic volatilities. In spite of this
flexibility, the design of a Libor model that can be calibrated in a feasible
way to a (in some sense) complete set of liquid market quotes (e.g. caps and
swaptions for different strikes and different maturities), remains a delicate
problem however. In its early version, the Libor model was usually driven by a
set of Brownian motions and equipped with some deterministic volatility
structure. These Libor models, termed \textit{market models,} where quite
popular because they allow for analytic cap(let) pricing and (approximate)
analytic swaption pricing via Black 76 type formulas. However, a main drawback
of these Libor market models is that they cannot match implied volatility
\textquotedblleft smile/skew\textquotedblright\ behavior observed in the cap
and swap markets. Moreover, these smile/skew effects became ever more
pronounced over the years.

For incorporating smile/skew behavior into the Libor model several proposals
have been made, for example, the Constant Elasticity of Variance (CEV) based
extension of the Libor market model by \cite{AA}, and the displaced diffusion
Libor market model by \cite{JR}.\ The implied volatility patterns produced by
these two approaches have the problem that they are of monotonic nature, so
only positive or negative skew effects can be imaged. Brigo and Mercurio
propose in \cite{BM} a local volatility model consistent with a mixture of
lognormal transition densities and some variations on this. One of the
problems in this approach is the rather complicated volatility structure
necessary for Monte Carlo simulation of the model in some fixed (e.g.
terminal) measure, and the limited flexibility for matching too pronounced
smile/skew market data. One further line of research on smile/skew explaining
Libor models concentrates on Libor models driven by compound Poisson processes
\cite{GK}, or even infinite activity L\'{e}vy processes \cite{EO}.
Particularly, in \cite{BS} a specifically structured jump driven Libor model
is developed that allows for feasible sequential calibration to cap
volatility-strike data for a whole system of maturities. Generally speaking,
however, Monte Carlo simulation of jump driven Libor models is rather
troublesome and expensive due to an unavoidable complicated drift term.
Recently, in \cite{PSS} an improvement is established in this respect, by
constructing L\'{e}vy approximations to this Libor drift. In the work of
\cite{WuZha} a Heston version of the Libor market model is proposed. In the
dynamics of this model, which is related to the models in \cite{P} and
\cite{ABR}, the volatility of each forward Libor $L_{i}$ (spanning over the
interval $[T_{i},T_{i+1}]$) contains a common stochastic volatility factor
$\sqrt{v}$ where $v$ is a Cox-Ingersoll-Ross type square-root process,
\emph{correlated} with Libor driving Brownian motions. Moreover, \cite{WuZha}
shows that their model has strong potential to produce smiles and skews (in
particular due to the correlation of $v$), and they present Fourier based
quasi analytic approximation methods for the pricing of caps and swaptions.
Therefore, in some sense the paper of \cite{WuZha} may be considered as a
first important step towards stochastic volatility Libor modeling.
Nonetheless, what is missing in this article and in most of the works
mentioned above is the assessment of the capability of the respective models
to be calibrated to a larger system of market quotes, including the cap(let)
volatility-strike (short capvola-strike) panels for a whole system of
maturities. In particular, it turned out that only one common volatility
factor as in the model of \cite{WuZha} may not be sufficient for matching a
larger set of cap volatility-strike panels that vary significantly over
different maturities. The reason is clear: A single stochastic volatility
factor determines a specific volatility-strike profile that may be consistent
with the market profile over one or some more maturities, but may not over a
complete tenor structure spanning twenty years for example. As a way out,
\cite{BMS} designed a multi-factor stochastic volatility model involving a
Brownian motion $W:=(W_{k})_{1\leq k<n},$ where the dimension of $W$ is equal
to the number of Libors, and each component is weighted with a (generally
different) square-root type stochastic factor $v_{k},$ and deterministic
loading factor $\beta_{ik},$ leading to a stochastic structure
\begin{equation}
\frac{dL_{i}}{L_{i}}=...dt+\sum_{k=i}^{n-1}\beta_{ik}\sqrt{v_{k}}dW_{k},\text{
\ \ }1\leq i<n, \label{u}%
\end{equation}
for the forward Libor $L_{i}$ (under some particular measure).~The technical
advantage of this approach is that, after standard freezing of the respective
Libors $L_{j}$ to $L_{j}(0)$ in the drift of the dynamics of $L_{i},$ a pure
affine Libor dynamics is produced, and as a consequence, caps and swaptions
can be priced quasi-analytically by a straightforward extension of the pricing
methods in \cite{WuZha}.\ On the other hand, the model of \cite{BMS} allows
for much greater flexibility with regard to calibration to a full system of
capvola-strike-maturity data. Of course the latter doesn't come as a surprise
since (\ref{u}) is in fact a generalization of the model in \cite{WuZha} (that
is retrieved by taking $v_{k}\equiv v$). Essentially, in \cite{BMS} the
volatility processes in (\ref{u}) are calibrated sequentially to the
capvola-strike data in the following way. One calibrates the process $v_{n-1}$
to the (last) vola-strike panel due to $L_{n-1}.$ Next one calibrates
$v_{n-2}$ to the vola-strike data involving $L_{n-2}$ with $v_{n-1}$ already
being identified, and one so works all the way back. After carrying out many
calibration tests with the model in \cite{BMS} it turned out that the
calibration works well as long as there are no big structural movements in the
capvola-strike patterns when going one step down from $T_{i}$ to $T_{i-1}.$
Indeed, in the particular case when a larger number of volatility processes
are already identified, say $v_{5},...,v_{40}$ with $n=41$ for an instance,
then a single additional volatility process $v_{4}$ may not be able to match a
panel at $T_{4}$ with a sudden strongly deviating vola-strike profile. In
fact, such breaks in the vola-strike patterns where quite typical during the
crisis. In this paper we present a new flexible multi-factor stochastic
volatility Libor model that resolves this problem and remains robust even in
more critical financial times.

\smallskip

The central theme of the present paper is a generalization of the Wu-Zhang
model in the following direction, i.e. we study processes
\begin{equation}
\frac{dL_{i}}{L_{i}}=...dt+\sqrt{v_{i}}\beta_{i}^{\top}dW,\text{ \ \ \ }1\leq
i<n, \label{uu}%
\end{equation}
where by taking $v_{i}\equiv v$ the Wu-Zhang model is retrieved again. In
contrast to the structure (\ref{u}), the danger of cumulative cementation of
the model in a backward recursive calibration is abandoned. Moreover, the
dimension of $W$ is not strongly restricted anymore to the number of Libors,
in order to render a recursive calibration as in (\ref{u}). However, several
technical issues have to be resolved. As a main point, even after standard
Libor freezing in the drift of the full stochastic differential equation (SDE)
corresponding to (\ref{uu}), we do not have an affine Libor model as in
\cite{WuZha} and \cite{BMS}\ anymore. That is, the Fourier based
quasi-analytical approximation for caps doesn't carry over directly. The same
complication shows up when one attempts to derive an approximate affine swap
market model from (\ref{uu}) in order to derive quasi-analytical (Fourier
based) swaption approximations. As a solution we will nevertheless construct
affine Libor approximations to (\ref{uu}) and affine swap rate approximations
connected with (\ref{uu}), that allow for quasi-analytical cap and swaption
pricing again. But, the price we have to pay is that these approximations are
typically (a bit) less accurate than the ones in the setting of \cite{WuZha}
and \cite{BMS}. Careful tests reveal that the approximation procedures
developed in this paper are accurate enough for our purposes however. The
bottom line and justification of our new approach is the following
\textquotedblleft philiosophical\textquotedblright\ point of view.\newline%
\ \newline\emph{A modeling package that contains only moderately accurate
procedures for calibrating to liquid market quotes (e.g. accuracy }$\sim
1\%$\emph{), but, which is able to achieve an adequate fitting error (e.g.
}$\sim3\%$\emph{ due to the }$1\%$\emph{ off pricing methods) in an efficient
way, is highly preferable in comparison to a modeling package that contains
very accurate pricing procedures for calibration (e.g. }$\leq0.2\%$\emph{
accurate), but, which is unable to achieve an adequate fitting error (e.g.
}$\sim10\%,$)\emph{ despite of the accurate pricing formulas. } \newline%
\ \newline Indeed, the former package achieves implicitly a fitting quality
with respect to the \textquotedblleft true model\textquotedblright\ of about
$4\%,$ while the latter package remains left at an unsatisfactory fit of
$\sim10.2\%.$ Further, for completeness, we extend the structure (\ref{uu})
with a standard Gaussian part and with displacement factors like in \cite{JR},
and consider the structure%
\begin{equation}
\frac{dL_{i}}{L_{i}+\alpha_{i}}=...dt+\sqrt{v_{i}}\beta_{i}^{\top}%
dW+\gamma_{i}^{\top}d\widehat{W},\text{ \ \ }1\leq i<n, \label{uuu}%
\end{equation}
where now $W$ and $\widehat{W}$ are independent standard Brownian motions,
$\gamma_{i}$ are deterministic factor loadings and $\alpha_{i}$ are
displacement constants for $1\leq i<n.$ From a technical point of view this
extension goes through without any difficulties, neither with regard to the
approximate pricing formulas, nor with regard to the new calibration
procedure. From a practical point of view it enlarges the flexibility of the
model, but in any particular case, the user can follow her taste and may set
$\gamma_{i}\equiv0,$ or $\alpha_{i}\equiv0,$ or both.

As a final introductory note we underline that the cap and swaption
approximation procedures proposed in this paper can be performed by (inverse)
Fast Fourier Transformation (FFT), and are thus rather fast. However, as an
alternative, the recently developed closed form approximation
for put/call options in a Heston model from \cite{Gob}, may be straightforwardly adapted to
closed form cap(let) and swaption pricing formulas in the context of the
(approximate) affine stochastic volatility Libor and swap rate model here
presented. Although we consider a detailed treatment here beyond scope, we
anticipate that the present stochastic volatility Libor model equipped with
these formulas might be considered an alternative to so called SABR Libor
models (cf. \cite{Mer}, \cite{Hag} and the references therein). While SABR
based models gain popularity because of their closed form approximations for
vanilla options based on (small time) heat kernel expansions, they are also
criticized somehow, for instance, because of their typically non mean
reverting stochastic volatilities.

\section{Recap of Wiener driven Libor modeling}

Let us fix a sequence of tenor dates $0=:T_{0}<T_{1}<\ldots<T_{n}$, called a
tenor structure. For each tenor date we consider a zero bond processes
$B_{i},\;i=1,\ldots,n,$ where each $B_{i}$ lives on the interval $[0,T_{i}]$
and ends with its face value $B_{i}(T_{i})=1.$ A system of forward Libors on
the given tenor structure is now defined by\
\begin{equation}
L_{i}(t):=\frac{1}{\delta_{i}}\left(  \frac{B_{i}(t)}{B_{i+1}(t)}-1\right)
,\quad0\leq t\leq T_{i},\;1\leq i<n, \label{Ld}%
\end{equation}
where the periods $\delta_{i}:=T_{i+1}-T_{i},\;i=1,\ldots,n-1,$ between two
consecutive tenor dates are termed day-count fractions. In fact, $L_{i}$ may
be seen as the annualized effective rate due to a forward rate agreement for
the period $[T_{i},T_{i+1}]$ contracted at time $t.$ According to this
agreement the interest $\delta_{i}L_{i}(T_{i})$ on the notional $1$ is to be
settled or payed at $T_{i+1}.$

\bigskip In this article we consider a framework where the Libor defining
zero-bonds $\left(  B_{i}\right)  _{i=1,\ldots,n}$ are adapted processes that
live on a filtered probability space $(\Omega,$ $(\mathcal{F}_{t})_{0\leq
t\leq T_{\infty}},$ $P),$ where $T_{\infty}\geq T_{n}$ is some finite time
horizon and the filtration $(\mathcal{F}_{t})$ is generated by some
$d$-dimensional standard Brownian motion $\mathcal{W}.$ Under some further
mild technical conditions (see \cite{J2} and \cite{J1} for details) there now
exists for each $i,$ $0\leq i<n,$ an $\mathbb{R}^{d}$-valued predictable
volatility process $\Gamma_{i}$ such that the Libor dynamics are given by%
\begin{equation}
\frac{dL_{i}}{L_{i}}=-\sum_{j=i+1}^{n-1}\frac{\delta_{j}L_{j}}{1+\delta
_{j}L_{j}}\Gamma_{i}^{\top}\Gamma_{j}dt+\Gamma_{i}^{\top}d\mathcal{W}%
^{(n)},\quad0\leq t\leq T_{i},\;1\leq i<n, \label{LM0}%
\end{equation}
where $\mathcal{W}^{(n)}$ is an equivalent standard Brownian motion under the
terminal num\'eraire measure $P_{n}$ induced by the terminal zero coupon bond
$B_{n}.$ That is, for all $j,$ $B_{j}/B_{n}$ are $P_{n}$-martingales. (In this
paper we do not dwell on issues concerning local versus true martingales.) For
some general fixed $i,$ $1\leq i<n$ we may consider instead the num\'eraire
measure $P_{i+1}$ induced by the bond $B_{j+1},$ and then for $1\leq j\leq i$
we obtain from (\ref{LM0}) the dynamics%
\begin{align}
\frac{dL_{j}}{L_{j}}  &  =\Gamma_{j}^{\top}\left(  -\sum_{k=j+1}^{n-1}%
\frac{\delta_{k}L_{k}}{1+\delta_{k}L_{k}}\Gamma_{k}dt+d\mathcal{W}%
^{(n)}\right) \nonumber\\
&  =-\sum_{k=j+1}^{i}\frac{\delta_{k}L_{k}}{1+\delta_{k}L_{k}}\Gamma_{j}%
^{\top}\Gamma_{k}dt+\Gamma_{j}^{\top}\left(  -\sum_{k=i+1}^{n-1}\frac
{\delta_{k}L_{k}}{1+\delta_{k}L_{k}}\Gamma_{k}dt+d\mathcal{W}^{(n)}\right)
\nonumber\\
&  =:-\sum_{k=j+1}^{i}\frac{\delta_{k}L_{k}}{1+\delta_{k}L_{k}}\Gamma
_{j}^{\top}\Gamma_{k}dt+\Gamma_{j}^{\top}d\mathcal{W}^{(i+1)},\text{ \ }1\leq
j\leq i. \label{LMi}%
\end{align}
Since due to (\ref{Ld}) $L_{i}$ is a martingale under $P_{i+1},$ it
automatically follows that $\mathcal{W}^{(i+1)}$ in (\ref{LMi}) is a standard
Brownian motion under the equivalent measure $P_{i+1}.$ Finally we note that
in the case where the $\Gamma_{j}$ are \textit{deterministic} we have the well
documented Libor {Market} Model (LMM) (see for example \cite{BM} and
\cite{Sch05} and the references therein).

\section{A new expiry-wise stochastic volatility model with displacement}

The general representation (\ref{LM0}) for the Libor dynamics will now be
structured towards a multi-factor stochastic volatility model of type
(\ref{uuu}). Let us take%
\begin{align}
\Gamma_{j}  &  =\left[
\begin{array}
[c]{c}%
\sqrt{v_{j}}\widetilde{\beta}_{j}\\
\widetilde{\gamma}_{j}\\
0
\end{array}
\right]  ,\text{ \ \ }\mathcal{W}^{(n)}\mathcal{=}\left[
\begin{array}
[c]{c}%
W^{(n)}\\
\widehat{W}^{(n)}\\
\overline{W}^{(n)}%
\end{array}
\right]  ,\text{ \ \ where}\nonumber\\
dv_{j}  &  =\kappa_{j}(\theta_{j}-v_{j})dt+\sqrt{v_{j}}\left(  \sigma
_{j}^{\top}d\widehat{W}^{(n)}+\overline{\sigma}_{j}^{\top}d\overline{W}%
^{(n)}\right)  ,\text{ \ \ }v_{j}(0)=\theta_{j}, \label{in1}%
\end{align}
where $W^{(n)},$ $\widehat{W}^{(n)},$ $\overline{W}^{(n)}$ are mutually
independent standard Brownian motions with dimensions $m,$ $\widehat{m},$ and,
$\overline{m},$ respectively, with $m+\widehat{m}+\overline{m}=d$. Further,
for $1\leq j<n,$ $\widetilde{\beta}_{j}$ and $\widetilde{\gamma}_{j}$ are
loading factors (in $\mathbb{R}^{m}$ and $\mathbb{R}^{\widehat{m}}$
respectively) to be specified below, and $v_{j}$ are square-root volatility
processes with parameters $\kappa_{j}$ (mean reversion speed), $\theta_{j}$
(mean reversion level), and $\sigma$ and $\overline{\sigma}$ are deterministic
\textquotedblleft vol of vol\textquotedblright\ factor loadings (in
$\mathbb{R}^{\widehat{m}}$ and $\mathbb{R}^{\overline{m}}$ respectively),
where (for convenience)
\begin{equation}
|\sigma_{j}|^{2}+|\overline{\sigma}_{j}|^{2}=:\varepsilon_{j}^{2}. \label{eps}%
\end{equation}
We thus get
\begin{align}
\frac{dL_{j}}{L_{j}}  &  =-\sum_{k=j+1}^{n-1}\frac{\delta_{k}L_{k}}%
{1+\delta_{k}L_{k}}\left(  \widetilde{\beta}_{j}^{\top}\widetilde{\beta}%
_{k}\sqrt{v_{j}v_{k}}+\widetilde{\gamma}_{j}^{\top}\widetilde{\gamma}%
_{k}\right)  dt\label{in}\\
&  +\sqrt{v_{j}}\widetilde{\beta}_{j}^{\top}dW^{(n)}+\widetilde{\gamma}%
_{j}^{\top}d\widehat{W}^{(n)},\quad\nonumber
\end{align}
together with (\ref{in1}). We next set
\begin{equation}
\widetilde{\gamma}_{j}=\frac{L_{j}+\alpha_{j}}{L_{j}}\gamma_{j},\text{
\ \ }\widetilde{\beta}_{j}=\frac{L_{j}+\alpha_{j}}{L_{j}}\beta_{j},
\label{til}%
\end{equation}
for deterministic loading factors $\beta_{j}$ and $\gamma_{j}$ (in
$\mathbb{R}^{m}$ and $\mathbb{R}^{\widehat{m}}$ respectively), and
displacement constants $\alpha_{j},$ $1\leq j<n,$ and we obtain from
(\ref{in}),%
\begin{align}
\frac{dL_{j}}{L_{j}+\alpha_{j}}  &  =-\sum_{k=j+1}^{n-1}\frac{\delta_{k}%
(L_{k}+\alpha_{k})}{1+\delta_{k}L_{k}}\left(  \beta_{j}^{\top}\beta_{k}%
\sqrt{v_{j}v_{k}}+\gamma_{j}^{\top}\gamma_{k}\right)  dt\nonumber\\
&  +\sqrt{v_{j}}\beta_{j}^{\top}dW^{(n)}+\gamma_{j}^{\top}d\widehat{W}%
^{(n)},\quad\label{ld}%
\end{align}
i.e. the new multi-factor stochastic volatility Libor model with displacement
and stochastic volatilities driven by (\ref{in1}). By applying It\^{o}'s
formula to the log-Libors, (\ref{ld}) becomes
\begin{align}
d\ln\left(  L_{j}+\alpha_{j}\right)   &  =-\frac{1}{2}\left\vert \gamma
_{j}\right\vert ^{2}dt-\frac{1}{2}v_{j}\left\vert \beta_{j}\right\vert
^{2}dt\nonumber\\
&  -\sum_{k=j+1}^{n-1}\frac{\delta_{k}(L_{k}+\alpha_{k})}{1+\delta_{k}L_{k}%
}\left(  \gamma_{j}^{\top}\gamma_{k}+\beta_{j}^{\top}\beta_{k}\sqrt{v_{j}%
v_{k}}\right)  dt\nonumber\\
&  +\sqrt{v_{j}}\beta_{j}^{\top}dW^{(n)}+\gamma_{j}^{\top}d\widehat{W}^{(n)}.
\label{logd}%
\end{align}
In Section~(\ref{capp}) we propose a pragmatic approximation that allows for
quasi-analytical caplet pricing in the context of to \eqref{logd}.

\subsection*{Instantaneous correlations}

For the mutual instantaneous Libor correlations we have%
\begin{align*}
\text{Cor}_{L_{j},L_{j^{\prime}}}  &  :=\frac{\frac{dL_{j}}{L_{j}}\cdot
\frac{dL_{j^{\prime}}}{L_{j^{\prime}}}}{\sqrt{\frac{dL_{j}}{L_{j}}\cdot
\frac{dL_{j}}{L_{j}}}\sqrt{\frac{dL_{j}}{L_{j}}\cdot\frac{dL_{j^{\prime}}%
}{L_{j^{\prime}}}}}=\frac{\widetilde{\gamma}_{j}^{\top}\widetilde{\gamma
}_{j^{\prime}}+\sqrt{v_{j}v_{j^{\prime}}}\widetilde{\beta}_{j}^{\top
}\widetilde{\beta}_{j^{\prime}}}{\sqrt{|\widetilde{\gamma}_{j}|^{2}%
+v_{j}|\widetilde{\beta}_{j}|^{2}}\sqrt{|\widetilde{\gamma}_{j^{\prime}}%
|^{2}+v_{j^{\prime}}|\widetilde{\beta}_{j^{\prime}}|^{2}}}\\
&  =\frac{\gamma_{j}^{\top}\gamma_{j^{\prime}}+\sqrt{v_{j}v_{j^{\prime}}}%
\beta_{j}^{\top}\beta_{j^{\prime}}}{\sqrt{|\gamma_{j}|^{2}+v_{j}|\beta
_{j}|^{2}}\sqrt{|\gamma_{j^{\prime}}|^{2}+v_{j^{\prime}}|\beta_{j^{\prime}%
}|^{2}}},
\end{align*}
which yields for $\gamma\equiv0,$ Cor$_{L_{j},L_{j^{\prime}}}=\frac{\beta
_{j}^{\top}\beta_{j^{\prime}}}{|\beta_{j}||\beta_{j^{\prime}}|},$ and for
$\beta\equiv0,$ Cor$_{L_{j},L_{j^{\prime}}}=\frac{\gamma_{j}^{\top}%
\gamma_{j^{\prime}}}{|\gamma_{j}||\gamma_{j^{\prime}}|}$ as usual. For the
instantaneous correlations between Libors and the stochastic volatilities we
have%
\begin{align}
\text{Cor}_{L_{j},v_{j^{\prime}}}  &  :=\frac{\frac{dL_{j}}{L_{j}}\cdot
dv_{j^{\prime}}}{\sqrt{\frac{dL_{j}}{L_{j}}\cdot\frac{dL_{j}}{L_{j}}}%
\sqrt{dv_{j^{\prime}}\cdot dv_{j^{\prime}}}}=\frac{\sqrt{v_{j}v_{j^{\prime}}%
}\widetilde{\beta}_{j}^{\top}\sigma_{j^{\prime}}}{\sqrt{|\widetilde{\gamma
}_{j}|^{2}+v_{j}|\widetilde{\beta}_{j}|^{2}}\sqrt{v_{j^{\prime}}%
(|\sigma_{j^{\prime}}|^{2}+|\overline{\sigma}_{j^{\prime}}|^{2})}}\nonumber\\
&  =\frac{\sqrt{v_{j}}\beta_{j}^{\top}\sigma_{j^{\prime}}}{\sqrt{|\gamma
_{j}|^{2}+v_{j}|\beta_{j}|^{2}}\varepsilon_{j^{\prime}}}. \label{ls}%
\end{align}
For $\gamma\equiv0$ we thus obtain%
\[
\text{Cor}_{L_{j},v_{j^{\prime}}}=\frac{\beta_{j}^{\top}\sigma_{j^{\prime}}%
}{|\beta_{j}|\varepsilon_{j^{\prime}}}.
\]
For the mutual instantaneous correlations between the stochastic volatilities
we get%
\[
\text{Cor}_{v_{j},v_{j^{\prime}}}:=\frac{dv_{j}\cdot dv_{j^{\prime}}}%
{\sqrt{dv_{j}\cdot dv_{j}}\sqrt{dv_{j^{\prime}}\cdot dv_{j^{\prime}}}}%
=\frac{\sigma_{j}^{\top}\sigma_{j^{\prime}}+\overline{\sigma}_{j}^{\top
}\overline{\sigma}_{j^{\prime}}}{\varepsilon_{j}\varepsilon_{j^{\prime}}}.
\]

\subsection{Discussion of the Wu-Zhang model as a special case}

\label{ex1} Let us take as a special case $\gamma\equiv0,$ $\alpha_{j}%
\equiv0,$ $\kappa_{j}\equiv\kappa,$ $\theta_{j}\equiv\theta,$ and for some
fixed unit vectors $e_{vol}\in\mathbb{R}^{m},$ $\overline{e}_{vol}%
\in\mathbb{R}^{\overline{m}},$ $\sigma_{j}\equiv\varepsilon\rho e_{vol},$ $,$
$\overline{\sigma}_{j}\equiv\varepsilon\sqrt{1-\rho^{2}}\overline{e}_{vol}$
where $\rho$ is a fixed correlation constant, $-1\leq\rho\leq1.$ We now are in
the setting of Wu-Zhang \cite{WuZha}, since all volatility processes coincide,
i.e. $v_{j^{\prime}}\equiv v,$ and (\ref{ls}) becomes
\begin{equation}
\text{Cor}_{L_{j},v_{j^{\prime}}}=\text{Cor}_{L_{j},v}=\rho e_{j}^{\top
}e_{vol}, \label{WZc}%
\end{equation}
where $\beta_{j} \equiv |\beta_{j}|e_{j}$ with $e_{j}\in\mathbb{R}^{m}.$ We note
that (\ref{WZc}) reflects a short coming of the Wu-Zhang model. The
instantaneous correlations between the Libor $L_{j}$ and the common stochastic
volatility factor may not be chosen for each $j$ as freely as somehow eqn (2.9) from
\cite{WuZha} suggests, and we have $\left\vert
\text{Cor}_{L_{j},v}\right\vert \leq\left\vert e_{j}^{\top}e_{vol}\right\vert
$ in particular! From another point of view, for realistic uniform skew behavior one
needs Cor$_{L_{j},v}<0$ for all $j,$ so that $e_{j}^{\top}e_{vol}$ has to have
at least a fixed sign and may not become too small for all $j.$ This in turn
implies severe restrictions on the mutual Libor correlation structure which is
usually taken to be an input.

As an intermediate extension of the Wu-Zhang model above we may consider the
case $\gamma\equiv0,$ $\alpha_{j}\equiv0,$ and then for some unit vectors
$e_{vol}\in\mathbb{R}^{m},$ $\overline{e}_{vol}\in\mathbb{R}^{\overline{m}},$
we take $\sigma_{j}\equiv\varepsilon_{j}\rho_{j}e_{vol},$ $,$ $\overline
{\sigma}_{j}\equiv\varepsilon_{j}\sqrt{1-\rho_{j}^{2}}\overline{e}_{vol}$
where $\rho_{j}$ are fixed correlation constants, $-1\leq\rho_{j}\leq1,$
depending on $j,$ and mean reverting speed and level may depend on $j$ also.
We then have
\[
\text{Cor}_{L_{j},v_{j}}=\rho_{j}e_{j}^{T}\overline{e}_{vol},
\]
hence for each particular $j,$ any correlation dominated by $\left\vert
e_{j}^{\top}e_{vol}\right\vert $ may be attained. Furthermore, as a main
feature of the multi-factor model (\ref{in1})-(\ref{logd}), we may have full
flexibility regarding the correlations (\ref{WZc}), by the structure given in
Section~\ref{struct}.

\begin{remark}
If $\alpha_{j}\equiv0,$ a Libor market model is retrieved by taking $\beta
_{j}\equiv0,$ or by taking $v_{j}(0)=\theta_{j}\equiv1,$ $\sigma_{j}%
\equiv\overline{\sigma}_{j}\equiv0.$ A further reason for including the LMM
term $\gamma_{j}^{\top}d\widehat{W}$ in the Libor noise might be to have some
extra freedom for calibrating to swaptions due to the fact that caplet prices
only depend on $\left\vert \gamma_{j}\right\vert .$
\end{remark}

\section{Approximate caplet pricing and calibration}

\label{capp}

For quasi-analytical caplet pricing we will construct an (approximate)
characteristic function of $L_{j}$ under $P_{j+1}.$ Let us write (\ref{ld})
as
\begin{align*}
\frac{dL_{j}}{L_{j}+\alpha_{j}}  &  =\sqrt{v_{j}}\beta_{j}^{\top}\left[
dW^{(n)}-\sum_{k=j+1}^{n-1}\frac{\delta_{k}(L_{k}+\alpha_{k})}{1+\delta
_{k}L_{k}}\beta_{k}\sqrt{v_{k}}dt\right] \\
&  +\gamma_{j}^{\top}\left[  d\widehat{W}^{(n)}-\sum_{k=j+1}^{n-1}\frac
{\delta_{k}(L_{k}+\alpha_{k})}{1+\delta_{k}L_{k}}\gamma_{k}dt\right] \\
&  =:\sqrt{v_{j}}\beta_{j}^{\top}dW^{(j+1)}+\gamma_{j}^{\top}d\widehat
{W}^{(j+1)}.\quad
\end{align*}
Since $L_{j}$ is a martingale under $P_{j+1},$ we necessarily have that
$dW^{(j+1)}$ and $d\widehat{W}^{(j+1)}$ are standard Brownian motions under
$P_{j+1}.$ Since the covariation processes $\langle\overline{W}^{(n)}%
,B_{j}\rangle\equiv0$ for all $j,$ it follows that $d\overline{W}%
^{(j+1)}=d\overline{W}^{(n)}$\ for all $j$ (cf. \cite{WuZha} and \cite{BMS}).
The dynamics of the stochastic volatility process $v_{j}$ under $P_{j+1}$ can
thus be written as%

\begin{gather*}
dv_{j}=\kappa_{j}(\theta_{j}-v_{j})dt+\sqrt{v_{j}}\overline{\sigma}_{j}^{\top
}d\overline{W}^{(j+1)}\\
+\sqrt{v_{j}}\sigma_{j}^{\top}\left[  dW^{(j+1)}+\sum_{k=j+1}^{n-1}%
\frac{\delta_{k}(L_{k}+\alpha_{k})}{1+\delta_{k}L_{k}}\beta_{k}\sqrt{v_{k}%
}dt\right] \\
=\underset{(\ast)}{\underbrace{\left(  \kappa_{j}(\theta_{j}-v_{j}%
)+\sum_{k=j+1}^{n-1}\frac{\delta_{k}(L_{k}+\alpha_{k})}{1+\delta_{k}L_{k}%
}\sigma_{j}^{\top}\beta_{k}\sqrt{v_{j}v_{k}}\right)  }}dt\\
+\sqrt{v_{j}}\left(  \sigma_{j}^{\top}dW^{(j+1)}+\overline{\sigma}_{j}^{\top
}d\overline{W}^{(j+1)}\right)  .
\end{gather*}
Thus, in order to obtain approximate affine dynamics for $v_{j}$ it is enough
to approximate $(\ast)$ with an expression that is affine in $v_{j}.$ Let us
therefore consider the pragmatic approximation%
\begin{equation}
\sqrt{v_{j}v_{k}}=\sqrt{v_{j}\frac{v_{k}Ev_{j}}{Ev_{j}}}\approx\sqrt
{v_{j}\frac{v_{j}Ev_{k}}{Ev_{j}}}\approx v_{j}\sqrt{\frac{\theta_{k}}%
{\theta_{j}}} \label{av}%
\end{equation}
(note that $E\,v_{k}=\theta_{k}$ due to the initial condition in (\ref{in1})).
In the Wu-Zhang setting we have $v_{j}\equiv v$ and thus, strict equality in
(\ref{av}) appears. Combining (\ref{av}) and usual freezing of Libors in
$(\ast)$ then leads to the following approximate volatility dynamics,
\begin{gather*}
dv_{j}\approx\kappa_{j}\theta_{j}dt+\left(  -\kappa_{j}+\sum_{k=j+1}%
^{n-1}\sqrt{\frac{\theta_{k}}{\theta_{j}}}\left[  \frac{\delta_{k}%
(L_{k}+\alpha_{k})}{1+\delta_{k}L_{k}}\right]  (0)\sigma_{j}^{\top}\beta
_{k}\right)  v_{j}dt\\
+\sqrt{v_{j}}\left(  \sigma_{j}^{\top}dW^{(j+1)}+\overline{\sigma}_{j}^{\top
}d\overline{W}^{(j+1)}\right)  .
\end{gather*}
With%

\begin{align}
\kappa_{j}^{(j+1)}  &  =\kappa_{j}-\sum_{k=j+1}^{n-1}\sqrt{\frac{\theta_{k}%
}{\theta_{j}}}\left[  \frac{\delta_{k}(L_{k}+\alpha_{k})}{1+\delta_{k}L_{k}%
}\right]  (0)\sigma_{j}^{\top}\beta_{k}\nonumber\\
\theta_{j}^{(j+1)}  &  =\frac{\kappa_{j}\theta_{j}}{\kappa_{j}^{(j+1)}}
\label{jp1}%
\end{align}
we thus obtain from (\ref{logd}) the approximative system%
\begin{gather}
d\ln\left(  L_{j}+\alpha_{j}\right)  =-\frac{1}{2}\left\vert \gamma
_{j}\right\vert ^{2}dt-\frac{1}{2}v_{j}\left\vert \beta_{j}\right\vert
^{2}dt+\sqrt{v_{j}}\beta_{j}^{\top}dW^{(j+1)}+\gamma_{j}^{\top}d\widehat
{W}^{(j+1)},\quad\label{X}\\
dv_{j}=\kappa_{j}^{(j+1)}\left(  \theta_{j}^{(j+1)}-v_{j}\right)
dt+\sqrt{v_{j}}\left(  \sigma_{j}^{\top}dW^{(j+1)}+\overline{\sigma}_{j}%
^{\top}d\overline{W}^{(j+1)}\right)  ,\text{ \ \ }v_{j}(0)=\theta
_{j}.\nonumber
\end{gather}
Now the main point is that, if moreover $\beta_{j},$ $\sigma_{j},$ and
$\overline{\sigma}_{j}$ are constant in time (piece-wise constant would be
enough in fact), (\ref{X}) is an affine structure that allows for Fourier
based (approximate) caplet pricing.

\subsection{Caplet pricing via characteristic function}

\label{cp1}

In general the price of a $T_{j}$-caplet with strike $K$ is given by
\begin{align*}
C_{j}(K)  &  =\delta_{j}B_{j+1}(0)E_{j+1}(L_{j}(T_{j})-K)^{+}\\
&  =B_{j+1}(0)\delta_{j}E_{j+1}(L_{j}(T_{j})+\alpha_{j}-\left(  K+\alpha
_{j}\right)  )^{+}\\
&  =B_{j+1}(0)\delta_{j}E_{j+1}(\left(  L_{j}(0)+\alpha_{j}\right)
e^{\ln\frac{L_{j}(T_{j})+\alpha_{j}}{L_{j}(0)+\alpha_{j}}}-\left(
K+\alpha_{j}\right)  )^{+}\\
&  =:B_{j+1}(0)\delta_{j}E_{j+1}(L_{j}^{disp}(0)e^{\ln\frac{L_{j}^{disp}%
(T_{j})}{L_{j}^{disp}(0)}}-K_{j}^{disp})^{+}.
\end{align*}
We may thus apply the Carr-Madan Fourier pricing method (outlined in the next
subsection) for caplets using
\[
\varphi_{j+1}^{disp},\text{ \ }\widehat{L}_{j}^{disp}(0),\text{ \ }%
K_{j}^{disp},
\]
where the characteristic function%
\begin{equation}
\varphi_{j+1}^{disp}(z\,;v):=E_{j+1}\left[  \left.  e^{\mathfrak{i}z\ln
\frac{L_{j}^{disp}(T_{j})}{L_{j}^{disp}(0)}}\right\vert v_{j}(0)=v\right]
\label{car}%
\end{equation}
may be obtained as follows. Let us abbreviate for fixed $j,$ $X^{0,x,v}%
(t):=\ln L_{j}^{disp}(t)=\ln\left(  L_{j}(t)+\alpha_{j}\right)  $ with
$X^{0,x,v}(0)=\ln L_{j}^{disp}(0)=\ln\left(  L_{j}(0)+\alpha_{j}\right)  =:x,$
and $V^{0,x,v}(t):=v_{j}(t)$ with $V^{0,x,v}(0)=v_{j}(0)=:v.$ Then by
(\ref{X}) (using (\ref{eps})), the generator of the vector process $(X,V)$ is
given by%
\begin{align*}
A  &  :=A_{x,v}:=\left(  -\frac{1}{2}\left\vert \gamma_{j}\right\vert
^{2}-\frac{1}{2}v\left\vert \beta_{j}\right\vert ^{2}\right)  \frac{\partial
}{\partial x}+\kappa_{j}^{(j+1)}\left(  \theta_{j}^{(j+1)}-v\right)
\frac{\partial}{\partial v}\\
&  +\frac{1}{2}\left(  \left\vert \gamma_{j}\right\vert ^{2}+v\left\vert
\beta_{j}\right\vert ^{2}\right)  \frac{\partial^{2}}{\partial x^{2}}%
+v\sigma_{j}^{\top}\beta_{j}\frac{\partial^{2}}{\partial x\partial v}+\frac
{1}{2}\varepsilon_{j}^{2}v\frac{\partial^{2}}{\partial v^{2}}.
\end{align*}
Let $\widehat{p}\left(  z,z^{\prime}\,;t,x,v\right)  $ satisfy the Cauchy
initial value problem
\begin{equation}
\frac{\partial\widehat{p}}{\partial t}=A\widehat{p},\text{ \ \ \ \ }%
\widehat{p}(z,z^{\prime}\,;0,x,v)=e^{\mathfrak{i}\left(  zx+z^{\prime
}v\right)  }. \label{Ca}%
\end{equation}
Then
\[
\widehat{p}\left(  z,z^{\prime}\,;t,x,v\right)  =Ee^{\mathfrak{i}\left(
zX^{0,x,v}(t)+z^{\prime}V^{0,x,v}(t)\right)  }.
\]
We are only interested in the solution for $z^{\prime}=0.$ Let us therefore
consider the ansatz%
\[
\widehat{p}\left(  z\,;t,x,v\right)  =\exp\left(  A(z;t)+B_{0}%
(z;t)x+B(z;t)v\right)
\]
with%
\begin{equation}
A(z;0)=0,\text{ \ \ \ }B_{0}(z;0)=\mathfrak{i}z,\text{ \ \ }B(z;0)=0.
\label{boun}%
\end{equation}
Substitution in (\ref{Ca}) yields,%
\begin{align*}
\left(  \frac{\partial A}{\partial t}+\frac{\partial B_{0}}{\partial t}%
x+\frac{\partial B}{\partial t}v\right)   &  =\left(  -\frac{1}{2}\left\vert
\gamma_{j}\right\vert ^{2}-\frac{1}{2}v\left\vert \beta_{j}\right\vert
^{2}\right)  B_{0}\\
&  +\kappa_{j}^{(j+1)}\left(  \theta_{j}^{(j+1)}-v\right)  B+\frac{1}%
{2}\left(  \left\vert \gamma_{j}\right\vert ^{2}+v\left\vert \beta
_{j}\right\vert ^{2}\right)  B_{0}^{2}\\
&  +v\sigma_{j}^{\top}\beta_{j}B_{0}B+\frac{1}{2}\varepsilon_{j}^{2}vB^{2},
\end{align*}
and we get the Riccati system%
\begin{align*}
\frac{\partial A}{\partial t}  &  =-\frac{1}{2}\left\vert \gamma
_{j}\right\vert ^{2}B_{0}+\kappa_{j}^{(j+1)}\theta_{j}^{(j+1)}B+\frac{1}%
{2}\left\vert \gamma_{j}\right\vert ^{2}B_{0}^{2}\\
\frac{\partial B_{0}}{\partial t}  &  =0\\
\frac{\partial B}{\partial t}  &  =-\frac{1}{2}\left\vert \beta_{j}\right\vert
^{2}B_{0}-\kappa_{j}^{(j+1)}B+\frac{1}{2}\left\vert \beta_{j}\right\vert
^{2}B_{0}^{2}+\sigma_{j}^{\top}\beta_{j}B_{0}B+\frac{1}{2}\varepsilon_{j}%
^{2}B^{2}.
\end{align*}
Taking into account (\ref{boun}) we get%
\begin{align*}
\frac{\partial A}{\partial t}  &  =-\frac{1}{2}\left\vert \gamma
_{j}\right\vert ^{2}\left(  \mathfrak{i}z+z^{2}\right)  +\kappa_{j}%
^{(j+1)}\theta_{j}^{(j+1)}B\\
\frac{\partial B}{\partial t}  &  =-\frac{1}{2}\left\vert \beta_{j}\right\vert
^{2}\left(  \mathfrak{i}z+z^{2}\right)  -\left(  \kappa_{j}^{(j+1)}%
-\mathfrak{i}z\sigma_{j}^{\top}\beta_{j}\right)  B+\frac{1}{2}\varepsilon
_{j}^{2}B^{2}.
\end{align*}
It is well known (see \cite{H}) that this system can be explicitly solved, but
depending on the chosen branch of the complex logarithm one may have different
representations for its solution. We follow Lord and Kahl's representation due
to the principal branch, see \cite{LK}\footnote{In a personal communication,
Roger Lord confirmed a typo in the published version and so referred to the
preprint version.}, and obtain%
\[
B(z;t)=\frac{a_{j}+d_{j}}{\varepsilon_{j}^{2}}\frac{1-e^{d_{j}t}}%
{1-g_{j}e^{d_{j}t}}%
\]
and%
\[
A(z;t)=-\frac{1}{2}\left(  \mathfrak{i}z+z^{2}\right)  \int_{0}^{t}\left\vert
\gamma_{j}\right\vert ^{2}ds+\frac{\kappa_{j}^{(j+1)}\theta_{j}^{(j+1)}%
}{\varepsilon_{j}^{2}}\left\{  \left(  a_{j}-d_{j}\right)  t-2\ln\left[
\frac{e^{-d_{j}t}-g_{j}}{1-g_{j}}\right]  \right\}
\]
with%

\begin{align*}
a_{j}  &  =\kappa_{j}^{(j+1)}-\mathfrak{i}z\sigma_{j}^{\top}\beta_{j}\\
d_{j}  &  =\sqrt{a_{j}^{2}+\left\vert \beta_{j}\right\vert ^{2}\left(
\mathfrak{i}z+z^{2}\right)  \varepsilon_{j}^{2}}\\
g_{j}  &  =\frac{a_{j}+d_{j}}{a_{j}-d_{j}}.
\end{align*}
Resuming, by taking $t=T_{j}$ we get for (\ref{car}),%
\begin{align}
\varphi_{j+1}^{disp}(z\,;v)  &  =e^{-\mathfrak{i}z\ln L_{j}^{disp}(0)}%
\widehat{p}\left(  z\,;T_{j},\ln L_{j}^{disp}(0),v\right) \nonumber\\
&  =\exp\left(  \widetilde{A}(z;T_{j})+B(z;T_{j})v\right)  \exp\left(
-\frac{1}{2}\left(  \mathfrak{i}z+z^{2}\right)  \int_{0}^{T_{j}}\left\vert
\gamma_{j}\right\vert ^{2}ds\right)  \label{cf}%
\end{align}
with
\begin{align*}
B(z;T_{j})  &  =\frac{a_{j}+d_{j}}{\varepsilon_{j}^{2}}\frac{1-e^{d_{j}T_{j}}%
}{1-g_{j}e^{d_{j}T_{j}}},\text{ \ \ and}\\
\widetilde{A}(z;t)  &  :=\frac{\kappa_{j}^{(j+1)}\theta_{j}^{(j+1)}%
}{\varepsilon_{j}^{2}}\left\{  \left(  a_{j}-d_{j}\right)  T_{j}-2\ln\left[
\frac{e^{-d_{j}T_{j}}-g_{j}}{1-g_{j}}\right]  \right\}  .
\end{align*}

\subsection*{Carr \& Madan inversion formula}

Following Carr and Madan \cite{CM}, the $T_{j}$-caplet price is now obtained
by the inversion formula,%
\begin{align}
C_{j}(K)  &  =\delta_{j}B_{j+1}(0)(L_{j}^{disp}(0)-K_{j}^{disp})^{+}%
+\nonumber\\
&  \frac{\delta_{j}B_{j+1}(0)L_{j}^{disp}(0)}{2\pi}\int_{-\infty}^{\infty
}\frac{1-\varphi_{j+1}^{disp}(z-\mathfrak{i};\theta_{j})}{z(z-\mathfrak{i}%
)}e^{-\mathfrak{i}z\ln\frac{K_{j}^{disp}}{L_{j}^{disp}(0)}}dz, \label{CK}%
\end{align}
where $\varphi_{j+1}^{disp}$ is given by (\ref{cf}) and we recall that
$v_{j}(0)=\theta_{j}.$ The integrand in (\ref{CK}) decays with order $z^{-2}$
if $\left\vert z\right\vert \rightarrow\infty,$ which is rather slow from a
numerical point of view. It is therefore advantageous to modify the inversion
formula in the following way. Let $\varphi_{j+1}^{\mathcal{B},disp}$ be the
characteristic function (\ref{car}) due to the Black model,
\[
L_{j}^{disp}(T_{j})=L_{j}^{disp}(0)e^{-\frac{1}{2}\left(  \sigma^{B}\right)
^{2}T_{j}+\sigma^{B}\sqrt{T_{j}}\varsigma},\text{ \ \ }\varsigma\in N(0,1)
\]
in the measure $P_{j+1},$ with a certain suitably chosen volatility
$\sigma_{j}^{B}.$ We then have (cf. Black's 76 formula)
\[
E_{j+1}\left(  L_{j}^{disp}(T_{j})-K^{disp}\right)  ^{+}=\mathcal{B}%
(L_{j}^{disp}(0),T_{j},\sigma^{B},K^{disp}),
\]
where
\begin{align*}
\mathcal{B}(L,T,\sigma,K)  &  :=L\mathcal{N}\left(  d_{+}\right)
-K\mathcal{N}\left(  d_{-}\right)  ,\text{ \ \ with}\\
d_{\pm}  &  :=\frac{\ln\frac{L}{K}\pm\frac{1}{2}\sigma^{2}T}{\sigma\sqrt{T}%
},\text{ \ \ and}\\
\varphi_{j+1}^{\mathcal{B},disp}(z\,;v)  &  =\varphi_{j+1}^{\mathcal{B}%
,disp}(z)=E_{j+1}e^{\mathfrak{i}z\left(  -\frac{1}{2}\left(  \sigma
^{B}\right)  ^{2}T_{j}+\sigma^{B}\sqrt{T_{j}}\varsigma\right)  }\\
&  =e^{-\frac{1}{2}\left(  \sigma^{B}\right)  ^{2}T_{j}\left(  z^{2}%
+\mathfrak{i}z\right)  }.
\end{align*}
Now applying Carr and Madan's formula to the Black model yields
\begin{align}
C_{j}^{\mathcal{B}}(K)  &  :=\delta_{j}B_{j+1}(0)\mathcal{B}(L_{j}%
^{disp}(0),T_{j},\sigma^{B},K_{j}^{disp})=\delta_{j}B_{j+1}(0)(L_{j}%
^{disp}(0)-K_{j}^{disp})^{+}\label{CK1}\\
&  +~\frac{\delta_{j}B_{j+1}(0)L_{j}^{disp}(0)}{2\pi}\int_{-\infty}^{\infty
}\frac{1-\varphi_{j+1}^{\mathcal{B},disp}(z-\mathfrak{i})}{z(z-\mathfrak{i}%
)}e^{-\mathfrak{i}z\ln\frac{K_{j}^{disp}}{L_{j}^{disp}(0)}}dz,\nonumber
\end{align}
and by subtracting ( \ref{CK1} ) from (\ref{CK}) we get%
\begin{gather}
C_{j}(K)=C_{j}^{\mathcal{B}}(K)+\label{cpa}\\
\frac{\delta_{j}B_{j+1}(0)L_{j}^{disp}(0)}{2\pi}\int_{-\infty}^{\infty}%
\frac{\varphi_{j+1}^{\mathcal{B},disp}(z-\mathfrak{i};v)-\varphi_{j+1}%
^{disp}(z-\mathfrak{i};\theta_{j})}{z(z-\mathfrak{i})}e^{-\mathfrak{i}%
z\ln\frac{K_{j}^{disp}}{L_{j}^{disp}(0)}}dz.\nonumber
\end{gather}
The latter inversion formula is usually much more efficient since typically
the integrand decays much faster than in (\ref{CK}).

\subsection{Putting the caplet approximation to the test}\label{capp1}

We now test the accuracy of the Fourier based  caplet pricing method (\ref{cpa}) via the approximative characteristic function (\ref{cf}). In this respect we compare, for each particular $j,$ the simulation price of the ``true'' model (\ref{ld}) with the simulation price due to the model obtained by replacing each volatility dynamics $v_k,$ $k\neq j,$ with the process $v_j,$ yielding  a Wu-Zhang type approximation depending on $j$ in fact. In turn, the Fourier based $T_j$-caplet price approximation is known to be a very accurate approximation to the $j$-linked Wu-Zhang model, as already documented in \cite{WuZha}.

The initial Libor rates are stripped from a given spot interest rate curve (see Table~\ref{Tab0}). In the test model we drop  the Gaussian part, i.e. ${\gamma}_j\equiv 0,$ and also assume that no displacement is in force, i.e. $\alpha \equiv 0.$ We choose $\delta_j = T_{j+1} - T_j \equiv 1.0$ and we put \eqref{in1} and \eqref{ld}  according to Section~\ref{struct}, where
\begin{equation}
\beta_j=0.15 e_j, \quad  \text{such that} \qquad r_{ij}=e_i^\top e_j=e^{-0.073|T_i-T_j|}, \label{rij}
\end{equation}
and the other parameters  are given in Table \ref{Tab0}. The orthonormal vectors $e_j$ are obtained by a Cholesky decomposition of the correlation matrix $(r_{ij})$. The parameters for the stochastic volatility processes are taken to be representative for a typical calibration. In particular
they are chosen in such a way that the Feller condition $2\kappa\theta>\sigma^2$ is violated.
The mean reversion levels are uniformly set to $\theta_j \equiv 1.$
We compare caplet prices due to the ``true'' model and the approximative one, by Monte Carlo simulation based on $30,000$ simulated paths
(Table~\ref{Tab1}).

\begin{table}[ht]
\medskip%
\centering
\begin{tabular}
[c]{|c|c|c|c|c|c|}\hline
$j$	&$\rho_j$	&$\kappa_j$	&$\varepsilon_j$	&$B_j(0)$	&$L_j(0)$\\
\hline\hline
1 &-0.70 &4.00000000 &3.00000000 &0.971717  &0.0332468\\
2 &-0.70 &3.95918367 &2.97959184 &0.94045  &0.0257067\\
3 &-0.70 &3.91836735 &2.95918367 &0.91688  &0.0195338\\
4 &-0.70 &3.87755102 &2.93877551 &0.899313  &0.0235296\\
5 &-0.70 &3.83673469 &2.91836735 &0.878639  &0.0278511\\
6 &-0.70 &3.79591837 &2.89795918 &0.854831  &0.0258653\\
7 &-0.70 &3.75510204 &2.87755102 &0.833278  &0.02359\\
8 &-0.70 &3.71428571 &2.85714286 &0.814074  &0.0237439\\
9 &-0.70 &3.67346939 &2.83673469 &0.795193  &0.0240497\\
10 &-0.70 &3.63265306 &2.81632653 &0.776518  &0.023694\\
11 &-0.70 &3.59183673 &2.79591837 &0.758545  &0.0234799\\
12 &-0.70 &3.55102041 &2.77551020 &0.741143  &0.0236513\\
13 &-0.70 &3.51020408 &2.75510204 &0.724019  &0.0238636\\
14 &-0.70 &3.46938776 &2.73469388 &0.707144  &0.0240064\\
15 &-0.70 &3.42857143 &2.71428571 &0.690566  &0.0241881\\
16 &-0.70 &3.38775510 &2.69387755 &0.674257  &0.0244311\\
17 &-0.70 &3.34693878 &2.67346939 &0.658177  &0.0246647\\
18 &-0.70 &3.30612245 &2.65306122 &0.642334  &0.024855\\
19 &-0.70 &3.26530612 &2.63265306 &0.626756  &0.0249485\\
\hline
\end{tabular}
\caption{Parameters of the Libor model, present values and initial Libor rates, terminal bond  $B_{20}(0) = 0.6115.$}
\label{Tab0}
\end{table}


\medskip

The numerical results show that  \eqref{X} approximates very accurately the true model dynamics  \eqref{in1} and \eqref{ld}. Indeed, the absolute price deviations are of magnitudes within basis points, with a well behaved relative error for ITM (in-the-money) and ATM (at-the-money) contracts. The relative errors become somewhat larger for OTM contracts, but OTM (out-of-the-money) caplet prices are typically very low (close to worthlessness) so that relative errors stemming from approximation \eqref{av}, \eqref{X} are intrinsically unstable (for any ``good'' approximation in fact).

\begin{table}[ht]
\medskip%
\centering
\begin{tabular}
[c]{|c||c|c|c|c|c|}\hline
$T_j$	&Strike	&Price (SE)	&Approx. price (SE)	&Abs. error	&Rel. error\\
\hline\hline
	&0.000	&0.0245 (9.28e-05)	&0.0244 (9.00e-05)	&1.71e-04	&0.0069\\
	&0.005	&0.0201 (8.96e-05)	&0.0200 (8.68e-05)	&1.66e-04 &0.0082\\
	&0.010	&0.0158 (8.62e-05)	&0.0156 (8.34e-05)	&1.64e-04	&0.0104\\
5.0	&0.015	&0.0115 (8.12e-05)	&0.0113 (7.85e-05)	&1.75e-04	&0.0151\\
	&0.020	&0.0076 (7.25e-05)	&0.0075 (7.00e-05)	&1.97e-04	&0.0255\\
&0.025	&0.0045 (5.96e-05)	&0.0043 (5.72e-05)	&2.03e-04	&0.0445\\
	&0.030	&0.0023 (4.45e-05)	&0.0022 (4.20e-05)	&1.74e-04	&0.0729\\	
\hline\hline
	&0.000	&0.0179 (9.91e-05)	&0.0177 (9.45e-05)	&2.50e-04	&0.0139\\
	&0.005	&0.0141 (9.61e-05)	&0.0139 (9.15e-05)	&2.45e-04 &0.0173\\
	&0.010	&0.0105 (9.16e-05)	&0.0102 (8.72e-05)	&2.56e-04	&0.0243\\
11.0	&0.015	&0.0073 (8.36e-05)	&0.0070 (7.94e-05)	&2.74e-04	&0.0375\\
	&0.020	&0.0047 (7.24e-05)	&0.0045 (6.82e-05)	&2.73e-04	&0.0571\\
	&0.025	&0.0029 (5.97e-05)	&0.0027 (5.56e-05)	&2.45e-04	&0.0823\\
	&0.030	&0.0018 (4.85e-05)	&0.0016 (4.44e-05)	&1.99e-04	&0.109\\	
\hline\hline
	&0.000	&0.0168 (1.06e-04)	&0.0165 (1.00e-04)	&2.81e-04	&0.0166\\
	&0.005	&0.0134 (1.04e-04)	&0.0131 (9.86e-05)	&2.79e-04 &0.0208\\
	&0.010	&0.0101 (1.00e-04)	&0.0098 (9.48e-05)	&2.95e-04	&0.0290\\
15.0	&0.015	&0.0074 (9.29e-05)	&0.0070 (8.76e-05)	&3.14e-04	&0.0423\\
	&0.020	&0.0052 (8.31e-05)	&0.0049 (7.78e-05)	&3.14e-04	&0.0602\\
	&0.025	&0.0035 (7.22e-05)	&0.0033 (6.69e-05)	&2.92e-04	&0.0813\\
	&0.030	&0.0024 (6.14e-05)	&0.0021 (5.62e-05)	&2.53e-04	&0.1043\\	
\hline\hline
	&0.000	&0.0158 (1.03e-04)	&0.0155 (9.81e-05)	&2.74e-04	&0.0172\\
	&0.005	&0.0127 (1.03e-04)	&0.0124 (9.74e-05)	&2.77e-04 &0.0217\\
	&0.010	&0.0098 (1.00e-04)	&0.0095 (9.46e-05)	&2.98e-04	&0.0302\\
19.0	&0.015	&0.0074 (9.43e-05)	&0.0071 (8.88e-05)	&3.19e-04	&0.0430\\
	&0.020	&0.0055 (8.62e-05)	&0.0051 (8.08e-05)	&3.25e-04	&0.0509\\
	&0.025	&0.0040 (7.72e-05)	&0.0037 (7.17e-05)	&3.12e-04	&0.0773\\
	&0.030	&0.0029 (6.81e-05)	&0.0026 (6.26e-05)	&2.84e-04	&0.0963\\
\hline
\end{tabular}
\caption{Simulation results for caplets.}
\label{Tab1}
\end{table}

\subsection{Further structuring and calibration}

\label{struct}

As part of the model, we choose a fixed LMM part $\gamma_{j}$ of the Libor
structure. This part may be obtained from an LMM calibration, eventually
weighted with some factor for instance or, if enough flexibility is left for
our purposes, we may set $\gamma_{j}\equiv0.$ The loadings $\beta_{j}$ are
also assumed to be chosen in advance. We further take $\overline{m}=1$ in
(\ref{in1}), and for $\rho_{j},$ $-1\leq\rho_{j}\leq1,$ we take $\sigma
_{j}=:\varepsilon_{j}\rho_{j}e_{j},$ where $\beta_{j}=:|\beta_{j}|e_{j},$ and
so $\overline{\sigma}_{j}=:\sqrt{1-\rho_{j}^{2}}\varepsilon_{j}.$ Note that in
principle we have no restrictions on $\rho_{j}$ conferred to the Wu-Zhang case
(see Section~\ref{ex1}).\ Then (\ref{ls}) becomes%
\[
\text{Cor}_{L_{j},v_{j^{\prime}}}=\rho_{j^{\prime}}e_{j}^{\top}e_{j^{\prime}%
}=\rho_{j^{\prime}}r_{jj^{\prime}}%
\]
with $r_{jj^{\prime}}:=e_{j}^{\top}e_{j^{\prime}},$ and in particular we have
Cor$_{L_{j},v_{j}}=\rho_{j}.$ For the mutual correlations between the
volatility processes we so have%
\begin{equation}
\text{Cor}_{v_{j},v_{j^{\prime}}}=\rho_{j}\rho_{j^{\prime}}r_{jj^{\prime}%
}+\sqrt{1-\rho_{j}^{2}}\sqrt{1-\rho_{j^{\prime}}^{2}}. \label{cvv}%
\end{equation}
In any case the scalars $\kappa_{j},\theta_{j},\rho_{j},\varepsilon_{j},$ and
the loadings have to be time independent, in order to invoke standard
square-root volatility processes. In principle piecewise constant
$t\mapsto\beta_{j}(t)$ will allow for Fourier based caplet pricing later on,
but for simplicity we assume henceforth that the $\beta_{j}$ are also time independent.

\begin{remark}
\label{corv} In practice it turns out that the $\rho_{j}$ are negative overall
in order to produce a skew. Let us assume for simplicity that we could fit the
data with a uniform (negative or positive) $\rho.$ Then (\ref{cvv}) implies
Cor$_{v_{j},v_{j^{\prime}}}=1-\rho^{2}(1-r_{jj^{\prime}})$\ $\geq1-\rho^{2},$
assuming that mutual Libor correlations $r_{jj^{\prime}}$ are nonnegative.
This means that mutual correlations between volatility processes are typically
high ($\geq0.5$ for $\rho=0.7$), and even close to $1$ when $j^{\prime}$ is
close to $j.$
\end{remark}

\subsection{Calibration to caplet volatility-strike-maturity}

\label{cal}

\begin{figure}[pth]
\centering \includegraphics[width=6cm]{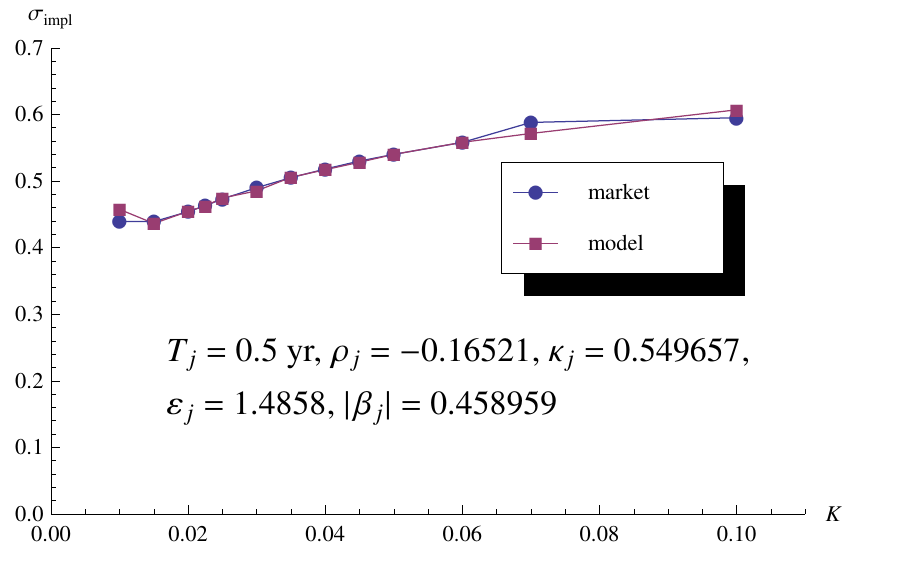}
\centering \includegraphics[width=6cm]{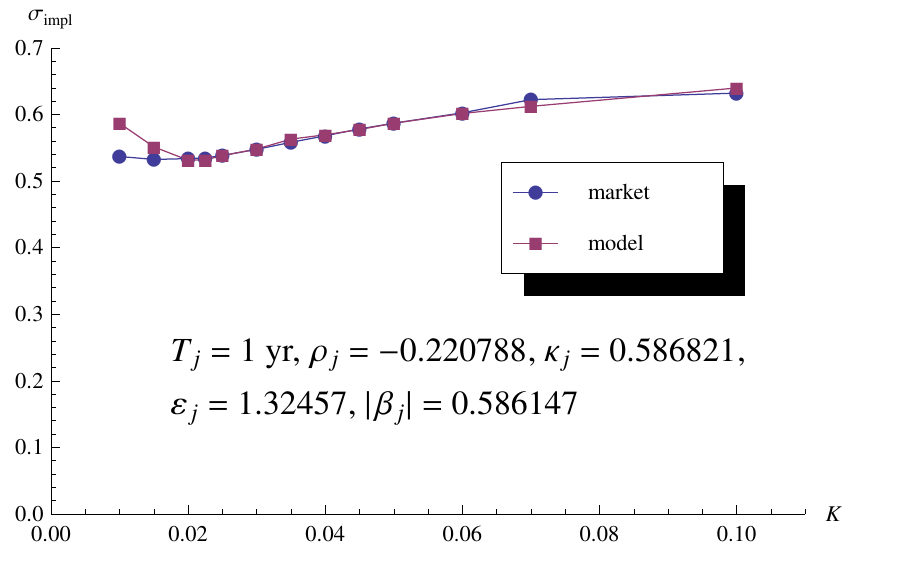} \newline%
\centering \includegraphics[width=6cm]{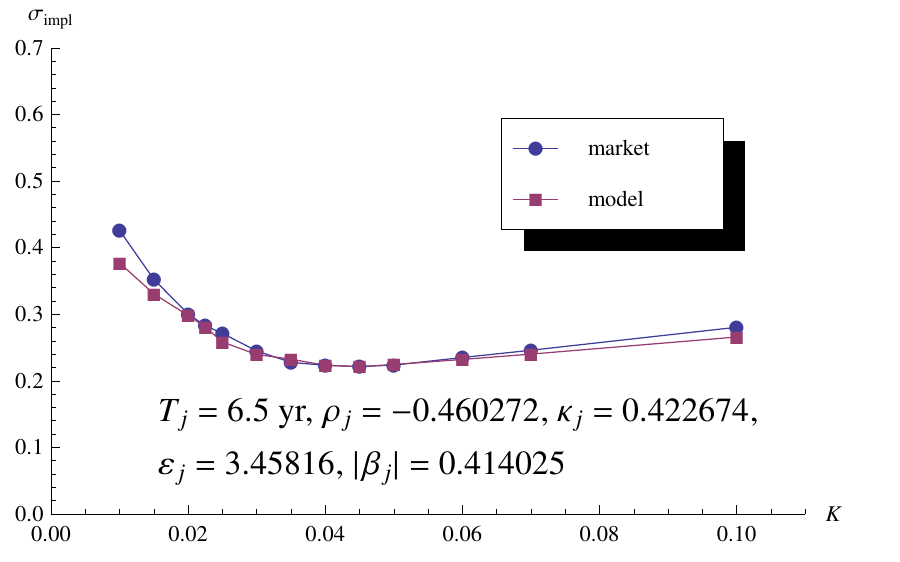}
\centering \includegraphics[width=6cm]{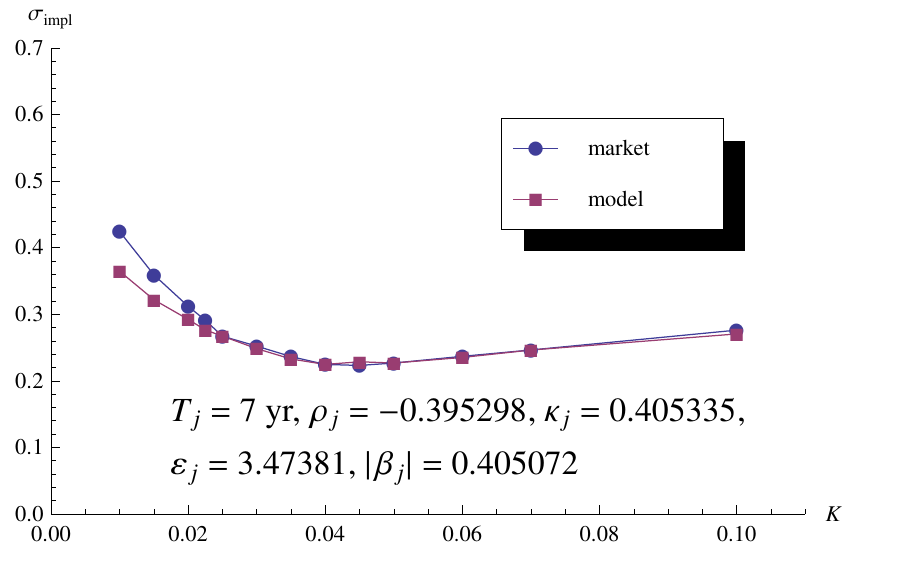} \newline%
\centering \includegraphics[width=6cm]{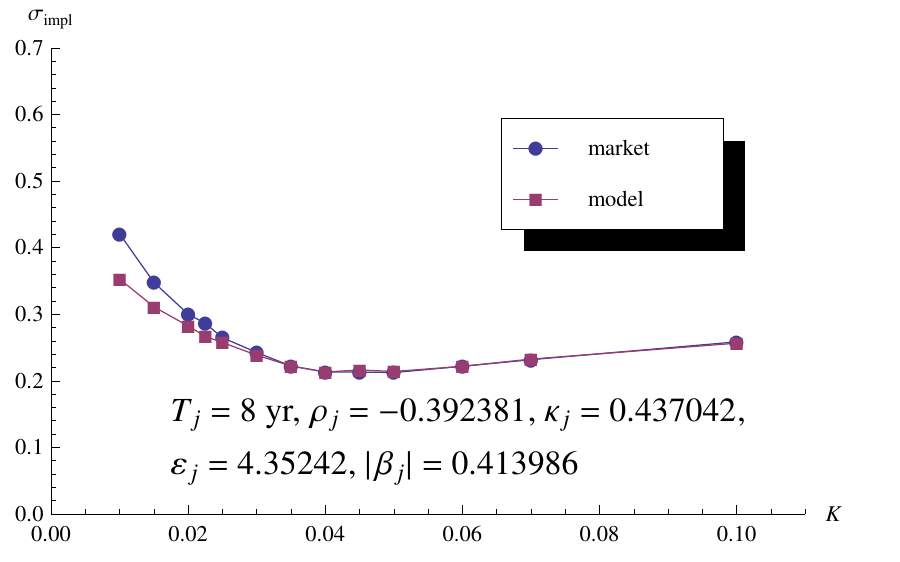}
\centering \includegraphics[width=6cm]{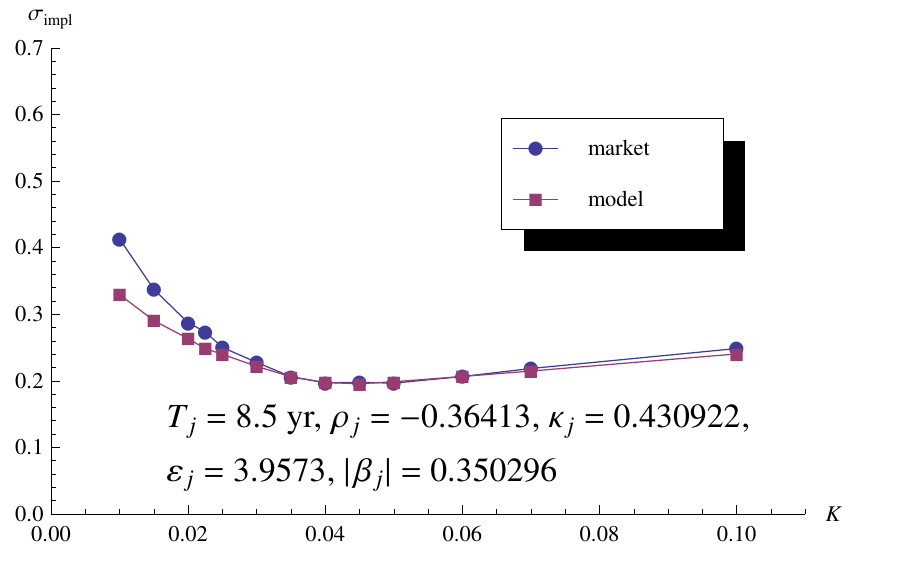} \newline%
\centering \includegraphics[width=6cm]{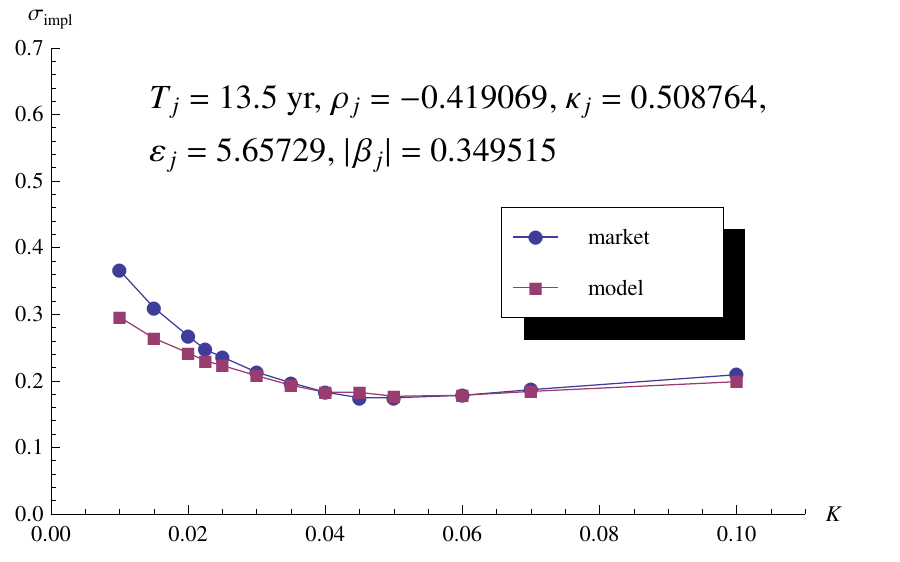}
\centering \includegraphics[width=6cm]{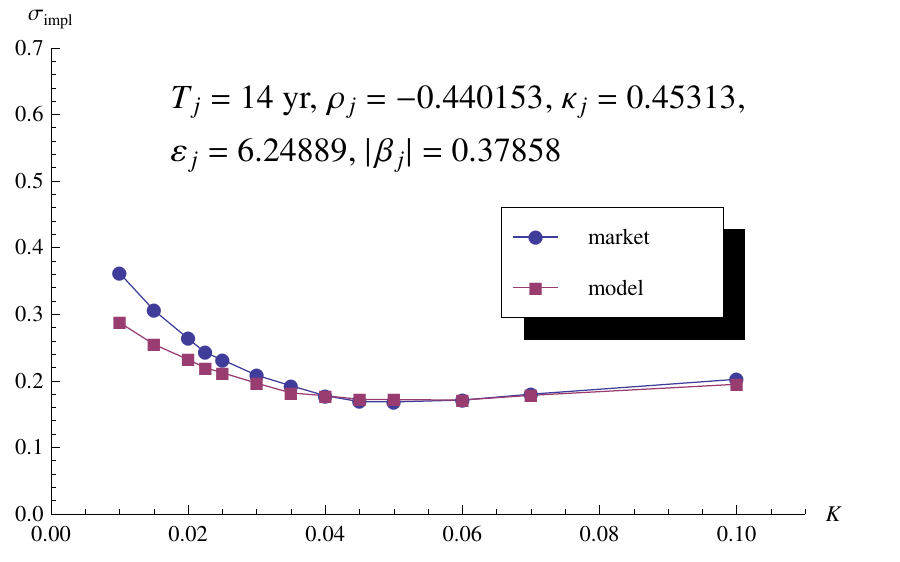}\newline%
\centering \includegraphics[width=6cm]{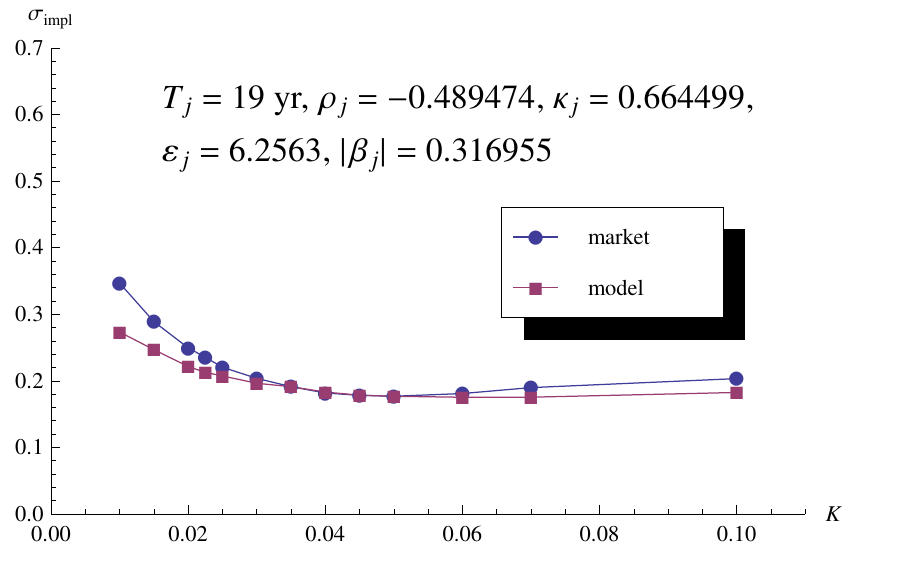}
\centering \includegraphics[width=6cm]{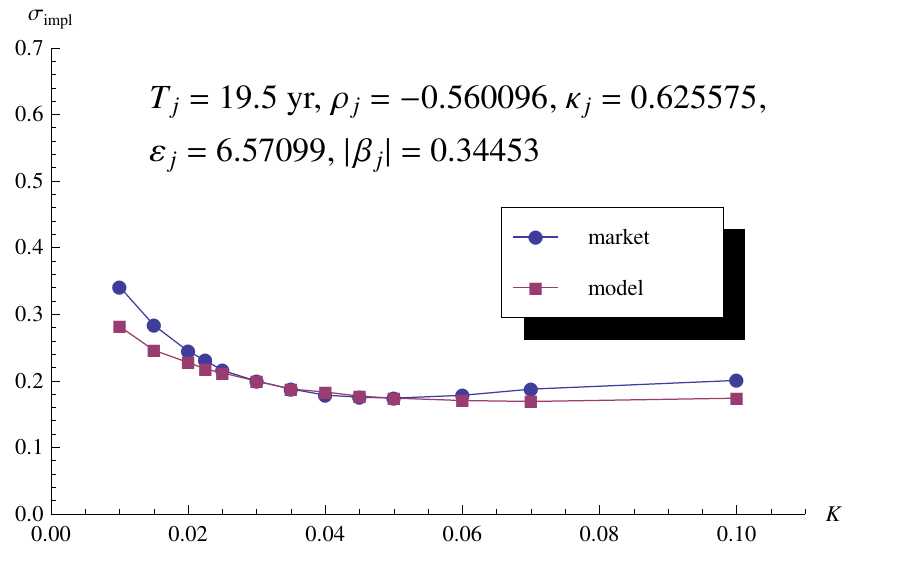}
\caption{Implied caplet volatilities due to market data vs. calibrated model}%
\label{LR_VR}%
\end{figure}

We will now illustrate a typical calibration test of the stochastic volatility
Libor model in its terminal measure to market cap-strike data. The test is
carried out for EurIBOR market data from September 20, 2010, based on a twenty
year semi-annual tenor structure. For simplicity, the displacements and the
Gaussian part where taken to be zero, i.e. $\alpha_{j}\equiv0,$ $\gamma
_{i}\equiv0,$ and as further input parameters we took $\theta_{i}\equiv1,$ and
$e_{i}$ from a Cholesky decomposition according to $e_{i}^{\top}e_{j}$ $=$
$r_{ij}$ $=$ $e^{-0.118|T_{i}-T_{j}|}$. For each maturity $T_{j},$ the
parameters
\[
\left\vert \beta_{j}\right\vert ,\text{ }\kappa_{j},\text{ }\varepsilon
_{j},\text{ }\rho_{j},
\]
where next calibrated to the caplet price-strike panel corresponding to
$T_{j},$ obtained from the market data. This calibration involves a minimum
search of a standard averaged relative error functional based on the FFT
pricing formula (\ref{CK}) due to the characteristic function (\ref{cf}). Each
trial $\kappa_{j}$ (which is restricted to $\kappa_{j}>0$) induces a
$\kappa_{j}^{(j+1)}$ and $\theta_{j}^{(j+1)}$ via (\ref{jp1}) (recall that
$\theta_{i}\equiv1$) which, together with $\rho_{j},$ are subsequently plugged
into (\ref{cf}). The implied volatility patterns due to the calibration as
well as the calibrated parameters are depicted in Figure~\ref{LR_VR}.
Concluding we may say that we obtained a satisfactory model fit with robustly
behaving parameters when moving from one maturity to the other. Optically the
fits for small strikes, hence deep ITM caplets may look a little bit
off overall. However, this is only appearance because our algorithm calibrates
to caplet \emph{prices}, while implied volatilities are badly conditioned for
deep ITM strikes.

\section{Swap rate dynamics and approximate swaption pricing}

\subsection{Swap contracts and dynamics under swap measures}

An interest rate swap is a contract to exchange a series of floating interest
payments in return for a series of fixed rate payments. Consider a series of
payment dates between $T_{p+1}$ and $T_{q},\,q>p.$ At each time $T_{j+1}%
,\,j=p,\ldots,q-1,$ the fixed leg of a (standard) swap pays $\delta_{j}K,$
whereas in return the floating leg pays $\delta_{j}L_{j}(T_{j})$ with
$L_{j}(T_{j})$ being the spot Libor rate. Consequently, the time $t-$value of
the interest rate swap (with $t\leq T_{p}$) is
\[
\sum_{j=p}^{q-1}\delta_{j}B_{j+1}(t)(L_{j}(t)-K).
\]
The swap rate $S_{p,q}(t)$ is defined to be the value of $K$ for which the
present value of the contract is zero. We thus have
\begin{equation}
S_{p,q}(t)=\frac{\sum_{j=p}^{q-1}\delta_{j}B_{j+1}(t)L_{j}(t)}{\sum
_{j=p}^{q-1}\delta_{j}B_{j+1}(t)}=\frac{B_{p}(t)-B_{q}(t)}{\sum_{j=p}%
^{q-1}\delta_{j}B_{j+1}(t)}. \label{SwapRate}%
\end{equation}
So $S_{p,q}$ is a martingale under the probability measure $P_{p,q}$, induced
by the annuity num\'eraire
\[
B_{p,q}(t):=\sum_{j=p}^{q-1}\delta_{j}B_{j+1}(t).
\]
From (\ref{ld}) it follows that
\begin{equation}
dS_{p,q}(t)=S_{p,q}(t)\Lambda_{p,q}^{\top}(t)d\mathcal{W}^{p,q}(t),
\label{SRSM1}%
\end{equation}
where $\mathcal{W}^{(p,q)}:=(W^{p,q},\widehat{W}^{p,q})$ is standard Brownian
motion under $P_{p,q},$ and where
\begin{equation}
\Lambda_{p,q}=\sum_{j=p}^{q-1}\frac{\delta_{j}\left(  L_{j}+\alpha_{j}\right)
}{1+\delta_{j}L_{j}}\left(  \sum_{l=j}^{q-1}w_{l}^{p,q}+\frac{B_{q}}%
{B_{p}-B_{q}}\right)  \left[
\begin{array}
[c]{c}%
\sqrt{v_{j}}\beta_{j}\\[0.3cm]%
\gamma_{j}%
\end{array}
\right]  , ~w_{l}^{p,q}:=\frac{\delta_{l}B_{l+1}}{B_{p,q}}. \label{sigpq}%
\end{equation}
The derivation hereof is given in Appendix~\ref{app}. We further have (see
Appendix~\ref{app}),%
\begin{equation}
d\mathcal{W}^{p,q}=d\mathcal{W}^{(n)}-dt\sum_{l=p}^{q-1}w_{l}^{p,q}%
\sum_{k=l+1}^{n-1}\frac{\delta_{k}\left(  L_{k}+\alpha_{k}\right)  }%
{1+\delta_{k}L_{k}}\left[
\begin{array}
[c]{c}%
\sqrt{v_{k}}\beta_{k}\\[0.3cm]%
\gamma_{k}%
\end{array}
\right]  .\text{ } \label{pqt}%
\end{equation}
By (\ref{SRSM1}) we thus get%
\begin{align}
d\ln S_{p,q}  &  =-\frac{1}{2}\frac{1}{S_{p,q}^{2}}d\langle S_{p,q}%
\rangle+\frac{dS_{p,q}}{S_{p,q}}\label{logSw}\\
&  =-\frac{1}{2}\left\vert \Lambda_{p,q}\right\vert ^{2}dt+\Lambda_{p,q}%
^{\top}d\mathcal{W}^{p,q},\nonumber
\end{align}
where by (\ref{sigpq}) we may write%
\begin{equation}
\Lambda_{p,q}=\sum_{j=p}^{q-1}\left[
\begin{array}
[c]{c}%
\sqrt{v_{j}}\beta_{j}\\[0.3cm]%
\gamma_{j}%
\end{array}
\right]  \frac{L_{j}+\alpha_{j}}{S_{p,q}}\mathfrak{\xi}_{j}^{p,q} \label{sa}%
\end{equation}
with%
\[
\mathfrak{\xi}_{j}^{p,q}:=\frac{\delta_{j}}{1+\delta_{j}L_{j}}\left(
\sum_{l=j}^{q-1}w_{l}^{p,q}\frac{B_{p}-B_{q}}{B_{p,q}}+\frac{B_{q}}{B_{p,q}%
}\right)  .
\]
(Cf. \cite{Sch05}, (1.35), and (1.38) so we have that $\mathfrak{\xi}%
_{j}^{p,q}(0)\approx w_{l}^{p,q}(0)$ with equality when the yield curve is
flat; hence the $\mathfrak{\xi}_{j}^{p,q}$ are approximate weights also.)

\subsection{Approximate affine swap rate dynamics}

\label{sec:approx_swap}

In order to approximate the swap rate process with a pure square-root
volatility process we introduce the process%
\begin{equation}
dv^{p,q}=\kappa^{p,q}(\theta^{p,q}-v^{p,q})dt+\sqrt{v^{p,q}}\left(
\sigma_{p,q}^{\top}dW^{(n)}+\overline{\sigma}_{p,q}^{\top}d\overline{W}%
^{(n)}\right)  ,\text{ \ \ }v^{p,q}(0)=\theta^{p,q} \label{avv}%
\end{equation}
with%
\begin{align}
\theta^{p,q}  &  :=\sum_{l=p}^{q-1}w_{l}^{p,q}(0)\theta_{l}\nonumber\\
\kappa^{p,q}  &  :=\sum_{l=p}^{q-1}w_{l}^{p,q}(0)\kappa_{l}\nonumber\\
\sigma_{p,q}  &  :=\sum_{l=p}^{q-1}w_{l}^{p,q}(0)\sigma_{l}\nonumber\\
\overline{\sigma}_{p,q}  &  :=\sum_{l=p}^{q-1}w_{l}^{p,q}(0)\overline{\sigma
}_{l}. \label{eq:setzungen}%
\end{align}
By replacing in (\ref{sa}) all volatility processes $v_{j}$ with the, in a
sense, averaged process $v^{p,q},$ and freezing Libors we arrive at the
approximation
\begin{align*}
\Lambda_{p,q}  &  \approx\sum_{j=p}^{q-1}\sum_{j=p}^{q-1}\left[
\begin{array}
[c]{c}%
\sqrt{v^{p,q}}\beta_{j}\\[0.3cm]%
\gamma_{j}%
\end{array}
\right]  \left[  \frac{L_{j}+\alpha_{j}}{S_{p,q}}\mathfrak{\xi}_{j}%
^{p,q}\right]  (0)\\
&  =\left[
\begin{array}
[c]{c}%
\sqrt{v^{p,q}}\beta_{p,q}\\[0.3cm]%
\gamma_{p,q}%
\end{array}
\right]  ,\text{ \ \ where}\\
\beta_{p,q}  &  :=\sum_{j=p}^{q-1}\beta_{j}\left[  \frac{L_{j}+\alpha_{j}%
}{S_{p,q}}\mathfrak{\xi}_{j}^{p,q}\right]  (0)\text{ \ and}\\
\gamma_{p,q}  &  :=\sum_{j=p}^{q-1}\gamma_{j}\left[  \frac{L_{j}+\alpha_{j}%
}{S_{p,q}}\mathfrak{\xi}_{j}^{p,q}\right]  (0)
\end{align*}
(note that $\sum_{j=p}^{q-1}\mathfrak{\xi}_{j}^{p,q}\left(  L_{j}+\alpha
_{j}\right)  /S_{p,q}\approx1$), hence yielding affine approximative swap rate
dynamics%
\begin{equation}
d\ln S_{p,q}=-\frac{1}{2}v^{p,q}\left\vert \beta_{pq}\right\vert ^{2}%
dt-\frac{1}{2}\left\vert \gamma_{p,q}\right\vert ^{2}dt+\sqrt{v^{p,q}}%
\beta_{p,q}^{\top}dW^{p,q}+\gamma_{p,q}^{\top}d\widehat{W}^{p,q}. \label{SRD}%
\end{equation}
For the (approximate) dynamics of $v^{p,q}$ under the annuity Brownian motions
we replace in (\ref{pqt}) the processes $v_{j}$ by their average $v^{p,q},$
and freeze the Libors as usual. \ From (\ref{avv}) we then obtain (as in
Section \ref{capp}, it follows again that $\overline{W}^{(n)}=\overline
{W}^{p,q}$ ),%
\begin{align}
dv^{p,q}  &  \approx\kappa^{p,q}(\theta^{p,q}-v^{p,q})dt+\sqrt{v^{p,q}%
}\overline{\sigma}_{p,q}^{\top}d\overline{W}^{(n)}\nonumber\\
&  +\sqrt{v^{p,q}}\sigma_{p,q}^{\top}\left(  dW^{p,q}+\sqrt{v^{p,q}}%
dt\sum_{l=p}^{q-1}\sum_{k=l+1}^{n-1}\left[  w_{l}^{p,q}\frac{\delta_{k}\left(
L_{k}+\alpha_{k}\right)  }{1+\delta_{k}L_{k}}\right]  (0)\,\beta_{k}\right)
.\nonumber
\end{align}
\ By setting
\begin{align*}
\widetilde{\kappa}^{p,q}  &  :=\kappa^{p,q}-\sum_{l=p}^{q-1}\left[
w_{l}^{p,q}\sum_{k=l+1}^{n-1}\frac{\delta_{k}\left(  L_{k}+\alpha_{k}\right)
}{1+\delta_{k}L_{k}}\right]  (0)\,\sigma_{p,q}^{\top}\beta_{k}\\
\widetilde{\theta}^{p,q}  &  =\frac{\kappa^{p,q}\theta^{p,q}}{\widetilde
{\kappa}^{p,q}},
\end{align*}
we thus have (in approximation)%
\begin{equation}
dv^{p,q}=\widetilde{\kappa}^{p,q}(\widetilde{\theta}^{p,q}-v^{p,q}%
)dt+\sqrt{v^{p,q}}\sigma_{p,q}^{\top}dW^{p,q}+\sqrt{v^{p,q}}\overline{\sigma
}_{p,q}^{\top}d\overline{W}^{p,q}. \label{avtil}%
\end{equation}

\subsection{Fourier based swaption pricing}

A (payer) swaption over the period $[T_{p},T_{q}]$ is the option to enter at
$T_{p}$ into a swap over the period $[T_{p},T_{q}]$ with strike $K.$ It
follows straightforwardly that the value at time $t=0$ is given by%
\begin{align}
\label{eq:payer_swaption}Swpn_{p,q}(K)=B_{p,q}(0)E_{p,q}\left[  \left(
S_{p,q}(T_{p})-K\right)  ^{+}\right]  .
\end{align}
Thus, after determining the characteristic function for $\ln\left[
S_{p,q}(T_{p})/S_{p,q}(0)\right]  $ we may price the option by the Carr-Madan
Fourier inversion method, just like we did for caplets in Section~\ref{cp1}.
Recalling the analysis from Section~\ref{cp1} it follows immediately that this
characteristic function is given by%
\begin{gather}
\varphi_{p,q}(z\,;v):=E_{p,q}\left[  \left.  e^{\mathfrak{i}z\ln\frac
{S_{p,q}(T_{p})}{S_{p,q}(0)}}\right\vert v_{p,q}(0)=v\right] \label{cfs}\\
\exp\left(  A_{p,q}(z;T_{p})+B_{p,q}(z;T_{p})v\right)  \exp\left(  -\frac
{1}{2}\left(  \mathfrak{i}z+z^{2}\right)  \int_{0}^{T_{p}}\left\vert
\gamma_{p,q}\right\vert ^{2}ds\right)  ,\nonumber
\end{gather}
where%

\[
B_{p,q}(z;T_{p})=\frac{a_{p,q}+d_{p,q}}{\varepsilon_{p,q}^{2}}\frac
{1-e^{d_{p,q}T_{p}}}{1-g_{p,q}e^{d_{p,q}T_{p}}}%
\]
and%
\[
A_{p,q}(z;T)=\frac{\widetilde{\kappa}^{p,q}\widetilde{\theta}^{p,q}%
}{\left\vert \sigma_{p,q}\right\vert ^{2}+\left\vert \overline{\sigma}%
_{p,q}\right\vert ^{2}}\left\{  \left(  a_{p,q}-d_{p,q}\right)  T_{p}%
-2\ln\left[  \frac{e^{-d_{p,q}T_{p}}-g_{p,q}}{1-g_{p,q}}\right]  \right\}
\]
with%

\begin{align*}
a_{p,q}  &  =\widetilde{\kappa}^{p,q}-\mathfrak{i}z\sigma_{p,q}^{\top}%
\beta_{p.q}\\
d_{p,q}  &  =\sqrt{a_{p,q}^{2}+\left\vert \beta_{p,q}\right\vert ^{2}\left(
\mathfrak{i}z+z^{2}\right)  \left(  \left\vert \sigma_{p,q}\right\vert
^{2}+\left\vert \overline{\sigma}_{p,q}\right\vert ^{2}\right)  }\\
g_{p,q}  &  =\frac{a_{p,q}+d_{p,q}}{a_{p,q}-d_{p,q}}.
\end{align*}
Based on \eqref{cfs} the (approximate) price of a swaption with maturity
$T_{p}$ and swaption leg $[T_{p},T_{q}]$ is given by
\begin{gather}
Swpn_{p,q}(K)=B_{p,q}(0)E_{p,q}\left[  \left(  S_{p,q}(T_{p})-K\right)
^{+}\right] \nonumber\\
\approx Swpn_{p,q}^{\mathcal{B}}(K)+\label{swpnF}\\
\frac{B_{p,q}(0)S_{p,q}(0)}{2\pi}\int_{-\infty}^{\infty}\frac{\varphi
_{p,q}^{\mathcal{B}}(z-\mathfrak{i};T_{p},\theta_{p,q})-\varphi_{p,q}%
(z-\mathfrak{i};T_{p},\theta_{p,q})}{z(z-\mathfrak{i})}e^{-\mathfrak{i}%
z\ln\frac{K}{S_{p,q}(0)}}dz\nonumber
\end{gather}
In \eqref{swpnF}, $\varphi_{p,q}^{\mathcal{B}}$ is the characteristic function
of a corresponding Black model,
\[
S_{p,q}(T_{p})=S_{p,q}(0)e^{-\frac{1}{2}\left(  \sigma_{p,q}^{B}\right)
^{2}T_{p}+\sigma_{p,q}^{B}\sqrt{T_{p}}\varsigma},\text{ \ \ }\varsigma\in
N(0,1),
\]
where $\sigma_{p,q}^{B}$ is a suitably chosen volatility, and
\[
Swpn_{p,q}^{\mathcal{B}}(K)=B_{p,q}(0)E_{p,q}\left(  S_{p,q}(T_{p})-K\right)
^{+}=B_{p,q}(0)\mathcal{B}(S_{p,q}(0),T_{p},\sigma_{p,q}^{B},K),
\]
is given by Black's formula (cf. \eqref{cpa}).

\subsection{Putting the swaption approximation to the test}
In the same spirit as we have tested the caplet price approximation in Section~\ref{capp1} we now test the above Fourier based swaption pricing method.
For each pair $(p,q),$ $1\le p<q\le n$ ($q\ne p+1$), we replace all volatility processes $v_j,$ $p\le j<q,$ with $v_{p,q}$ given by (\ref{avv}),
(\ref{eq:setzungen}) to obtain in fact a Wu-Zhang related swaption approximation model linked to this pair $(p,q).$ We then compare the simulated $(p,q)$-swaption price due to the ``true'' model (\ref{ld}) and the model with common stochastic volatility process (\ref{avv}). In turn, the latter price
can be accurately approximated by (\ref{swpnF}) as shown in  \cite{WuZha1}.
We base the numerical experiments on the same data set as in Section \ref{capp1}.

\smallskip

In detail, this means that for putting up the ``true'' and the approximate Libor model, the initial Libors are stripped from a given spot rate curve and their values are given in Table \ref{Tab0}, the Gaussian $\gamma$-part is deactivated by putting $\gamma_j\equiv 0$ and no displacement is in force by choosing $\alpha_j\equiv 0$. Moreover, the parametrization of the correlation structure from Section \ref{struct} is given by
\begin{align*}
&r_{ij} = \exp\big(-0.0553|T_i-T_j|\big) = e_i^\top e_j, ~ \beta_j = 0.15 e_j,
\end{align*}
with the orthonormal vectors $e_j$ resulting from a Cholesky decomposition of $(r_{ij})$ and $\delta_j = T_{j+1} - T_j \equiv 1.0$ and $\theta_j\equiv 1$ remain valid. All other simulation parameters, in particular the $\rho_j$'s, $\kappa_j$'s and $\varepsilon_j$'s can be found in Table \ref{Tab0} and we retain the diffusion coefficients
\begin{align*}
&\sigma_j = \rho_j \varepsilon_j e_j, ~ \overline{\sigma}_j = \sqrt{1-\rho_j^2}\varepsilon_j.
\end{align*}

\smallskip

\begin{table}[ht]
\medskip%
\centering
\begin{tabular}
[c]{|c||c|c|c|c|c|}\hline
$[T_p,T_q]$	&Strike	&Price (SE)	&Approx. price (SE)	&Abs. error	&Rel. error\\
\hline\hline
	& 0.000 &0.1640 (2.1e-04) &0.1637 (2.1e-04) &0.00032 &0.002\\  \relax
	& 0.005 &0.1302 (2.0e-04) &0.1299 (2.0e-04) &0.00032 &0.002\\ \relax
	& 0.010 &0.0964 (1.9e-04) &0.0961 (1.9e-04) &0.00031 &0.003\\ \relax
[2, 10]& 0.015 &0.0628 (1.8e-04) &0.0625 (1.8e-04) &0.00033 &0.005\\ \relax
	& 0.020 &0.0317 (1.5e-04) &0.0313 (1.5e-04) &0.00037	&0.011\\ \relax
	& 0.025 &0.0094 (9.0e-05) &0.0092 (9.0e-04) &0.00024	&0.026\\ \relax
	& 0.030 &0.0011 (3.0e-05) &0.0010 (2.9e-05) &0.00003 &0.030\\
\hline\hline
	&0.000 &0.1228 (2.3e-04) &0.1223 (2.3e-04) &0.00057 &0.004\\ \relax
	&0.005 &0.0981 (2.2e-04) &0.0975 (2.2e-04) &0.00055 &0.005\\ \relax
	& 0.010 &0.0734 (2.1e-04) &0.0728 (2.1e-04) &0.00055 &0.007\\ \relax
[4, 10]&0.015 &0.0493 (2.0e-04) &0.0488 (1.9e-04) &0.00057 &0.011\\ \relax
	&0.020 &0.0281 (1.6e-04) &0.0275 (1.6e-04) &0.00060	&0.021\\ \relax
	&0.025 &0.0127 (1.2e-04) &0.0122 (1.1e-04) &0.00049	&0.038\\ \relax
	&0.030 &0.0042 (7.1e-05) &0.0040 (6.9e-05) &0.00026 &0.060\\
\hline\hline
	& 0.000 &0.2877 (4.8e-04) &0.2866 (4.8e-04) &0.00110 &0.003\\ \relax
	& 0.005 &0.2288 (4.6e-04) &0.2277 (4.6e-04) &0.00107 &0.004\\ \relax
	& 0.010 &0.1699 (4.5e-04) &0.1689 (4.4e-04) &0.00104 &0.006\\ \relax
[4, 20]& 0.015 &0.1122 (4.2e-04) &0.1112 (4.2e-04) &0.00102 &0.009\\ \relax
	& 0.020 &0.0609 (3.5e-04) &0.0600 (3.5e-04) &0.00091	&0.015\\ \relax
	& 0.025 &0.0246 (2.4e-04) &0.0241 (2.4e-04) &0.00051	&0.020\\ \relax
	& 0.030 &0.0068 (1.2e-04) &0.0067 (1.2e-04) &0.00011 &0.016\\
\hline\hline
	& 0.000 &0.1653 (4.5e-04) &0.1638 (4.4e-04) &0.00149 &0.009\\ \relax
	& 0.005 &0.1311 (4.4e-04) &0.1297 (4.3e-04) &0.00146 &0.011\\ \relax
	& 0.010 &0.0976 (4.2e-04) &0.0961 (4.1e-04) &0.00146 &0.015\\ \relax
[10, 20]& 0.015 &0.0670 (3.9e-04) &0.0655 (3.8e-04) &0.00147 &0.021\\ \relax
	& 0.020 &0.0423 (3.3e-04) &0.0410 (3.3e-04) &0.00137	&0.032\\ \relax
	& 0.025 &0.0247 (2.7e-04) &0.0236 (2.6e-04) &0.00137	&0.045\\ \relax
	& 0.030 &0.0134 (2.0e-04) &0.0126 (1.9e-04) &0.00081 &0.060\\
\hline
\end{tabular}
\caption{Simulation results for payer swaptions.}
\label{TabB}
\end{table}

To gear towards the approximate Libor model, we perform the calculation of the weighted volatility parameters $\kappa^{p,q}$, $\theta^{p,q}$, $\sigma^{p,q}$ and $\bar{\sigma}^{p,q}$ according to \eqref{eq:setzungen}, where the frozen weights $w^{p,q}_l(0)$ are given in \eqref{sigpq}, so that the averaged approximate volatility process $v^{p,q}$ from \eqref{avv} can be simulated. This averaged stochastic volatility is then reinserted into the Libor dynamics \eqref{in}, i.e. $v^{p,q}$ virtually replaces each expiry-wise volatility $v_j, j=p,\ldots,q-1.$ The simulations are carried out using $30,000$ Monte Carlo paths.

\smallskip

We calculate ``true'' and approximate swaption prices for the payer swaption depicted in \eqref{eq:payer_swaption} for various strike levels and swap legs $[T_p, T_q]$. The results of our numerical experiments are depicted in Table \ref{TabB}.


\medskip

The simulation results show that for swaption pricing, the approximate Libor model under one weighted stochastic volatility $v^{p,q}$ gives a surprisingly good fit to the true model dynamics \eqref{in}, \eqref{in1}. Depending on the swap legs, absolute price deviations are in the range of basis points (for swaption maturing in two and four years) and in the range of ten basis points (for maturity ten years). Recalling that the approximation is somewhat strong as each expiry-wise volatility process $v_j, j=p,\ldots,q-1$ is replaced by one weighted volatility process $v^{p,q}$, the numerical results reveal however that we get reasonably well behaved approximations to the ``true'' model.


\section{Appendix}

\label{app} The derivation of the swap rate volatility (\ref{sigpq}) is
essentially given in \cite{Sch05}. But in order to match to the present
notation and to make reading more convenient, we now give a short recap. Let,
exclusively in this section, $\sigma_{j}$ denote the volatility of the bond
$B_{j},$ let $\mu_{j}$ be the drift of $B_{j},$ and $\lambda$ be the market
price of risk process with respect to the driving Brownian motion
$\mathcal{W=}(W,\widehat{W},\overline{W}).$ That is, in the objective measure
the zero bond dynamics are of the form%
\[
\frac{dB_{j}}{B_{j}}=\mu_{j}dt+\sigma_{j}^{\top}d\mathcal{W}\text{ \ \ with
\ \ }\mu_{j}=\sigma_{j}^{\top}\lambda,
\]
and where $\sigma_{j,k}=0$ for $m+\widehat{m}<k\leq m+\widehat{m}+\overline
{m}.$ Following \cite[p.17]{Sch05}, we may write%
\begin{align*}
dB_{p,q}  &  =\sum_{j=p}^{q-1}\delta_{j}dB_{j+1}=\cdot\cdot\cdot dt+\sum
_{j=p}^{q-1}\delta_{j}B_{j+1}\sigma_{j+1}^{\top}d\mathcal{W}\\
&  =\cdot\cdot\cdot dt+B_{p,q}\sum_{j=p}^{q-1}w_{j}^{p,q}\sigma_{j+1}^{\top
}d\mathcal{W}.
\end{align*}
We thus have by It\^o's formula for $p\leq r\leq q,$%
\begin{align*}
\frac{d(B_{r}/B_{p,q})}{B_{r}/B_{p,q}}  &  =\cdot\cdot\cdot dt+\left(
\sigma_{r}^{\top}-\sum_{j=p}^{q-1}w_{j}^{p,q}\sigma_{j+1}^{\top}\right)
d\mathcal{W}\\
&  =\left(  \sigma_{r}^{\top}-\sum_{j=p}^{q-1}w_{j}^{p,q}\sigma_{j+1}^{\top
}\right)  d\mathcal{W}^{p,q}%
\end{align*}
as $B_{r}/B_{p,q}$ is a $P_{p,q}$-martingale. We thus obtain
\begin{gather*}
dS_{p,q}=d\frac{B_{p}-B_{q}}{B_{p,q}}=\\
\left[  \frac{B_{p}}{B_{p,q}}\left(  \sigma_{p}^{\top}-\sum_{j=p}^{q-1}%
w_{j}^{p,q}\sigma_{j+1}^{\top}\right)  -\frac{B_{q}}{B_{p,q}}\left(
\sigma_{q}^{T}-\sum_{j=p}^{q-1}w_{j}^{p,q}\sigma_{j+1}^{\top}\right)  \right]
d\mathcal{W}^{p,q}\\
=\left[  \frac{B_{p}}{B_{p,q}}\sum_{j=p}^{q-1}w_{j}^{p,q}\left(  \sigma
_{p}^{\top}-\sigma_{j+1}^{\top}\right)  -\frac{B_{q}}{B_{p,q}}\sum_{j=p}%
^{q-1}w_{j}^{p,q}\left(  \sigma_{q}^{\top}-\sigma_{j+1}^{\top}\right)
\right]  d\mathcal{W}^{p,q}\\
=S_{p,q}\left[  \frac{B_{p}}{B_{p}-B_{q}}\sum_{j=p}^{q-1}w_{j}^{p,q}\left(
\sigma_{p}^{\top}-\sigma_{j+1}^{\top}\right)  -\frac{B_{q}}{B_{p}-B_{q}}%
\sum_{j=p}^{q-1}w_{j}^{p,q}\left(  \sigma_{q}^{\top}-\sigma_{j+1}^{\top
}\right)  \right]  d\mathcal{W}^{p,q}\\
=S_{p,q}\left[  \sum_{j=p}^{q-1}w_{j}^{p,q}\left(  \sigma_{p}^{\top}%
-\sigma_{j+1}^{\top}\right)  +\frac{B_{q}}{B_{p}-B_{q}}\sum_{j=p}^{q-1}%
w_{j}^{p,q}\left(  \sigma_{p}^{\top}-\sigma_{q}^{\top}\right)  \right]
d\mathcal{W}^{p,q}\\
=:S_{p,q}\Lambda_{p,q}^{\top}d\mathcal{W}^{p,q}.
\end{gather*}
Similar to (1.13) in \cite{Sch05} we get
\begin{align*}
\Lambda_{p,q}  &  =\sum_{j=p}^{q-1}w_{j}^{p,q}\left(  \sigma_{p}-\sigma
_{j+1}\right)  +\frac{B_{q}}{B_{p}-B_{q}}w_{j}^{p,q}\left(  \sigma_{p}%
-\sigma_{q}\right) \\
&  =\sum_{j=p}^{q-1}w_{j}^{p,q}\sum_{r=p}^{j}\left(  \sigma_{r}-\sigma
_{r+1}\right)  +\frac{B_{q}}{B_{p}-B_{q}}\sum_{r=p}^{q-1}\left(  \sigma
_{r}-\sigma_{r+1}\right) \\
&  =\sum_{r=p}^{q-1}\left(  \sigma_{r}-\sigma_{r+1}\right)  \left(  \sum
_{j=r}^{q-1}w_{j}^{p,q}+\frac{B_{q}}{B_{p}-B_{q}}\right) \\
&  =\sum_{r=p}^{q-1}\left[
\begin{array}
[c]{c}%
\sqrt{v_{r}}\beta_{r}\\[0.3cm]%
\gamma_{r}%
\end{array}
\right]  \frac{\delta_{r}\left(  L_{r}+\alpha_{r}\right)  }{1+\delta_{r}L_{r}%
}\left(  \sum_{j=r}^{q-1}w_{j}^{p,q}+\frac{B_{q}}{B_{p}-B_{q}}\right)  .
\end{align*}
Further, by (1.27) from \cite{Sch05}, it holds that
\[
d\mathcal{W}^{(n)}=d\mathcal{W}+(\lambda-\sigma_{n})dt,\text{ \ \ and
\ \ }d\mathcal{W}^{p,q}=\lambda dt-\sum_{l=p}^{q-1}w_{l}^{p,q}\sigma
_{l+1}dt+d\mathcal{W}.
\]
Therefore, we finally have%
\begin{align*}
d\mathcal{W}^{p,q}  &  =d\mathcal{W}^{(n)}+\sigma_{n}dt-\sum_{l=p}^{q-1}%
w_{l}^{p,q}\sigma_{l+1}dt\\
&  =d\mathcal{W}^{(n)}+dt\sum_{l=p}^{q-1}w_{l}^{p,q}\left(  \sigma_{n}%
-\sigma_{l+1}\right) \\
&  =d\mathcal{W}^{(n)}+dt\sum_{l=p}^{q-1}w_{l}^{p,q}\sum_{k=l+1}^{n-1}\left(
\sigma_{k+1}-\sigma_{k}\right) \\
&  =d\mathcal{W}^{(n)}-dt\sum_{l=p}^{q-1}w_{l}^{p,q}\sum_{k=l+1}^{n-1}%
\frac{\delta_{k}\left(  L_{k}+\alpha_{k}\right)  }{1+\delta_{k}L_{k}}\left[
\begin{array}
[c]{c}%
\sqrt{v_{k}}\beta_{k}\\[0.3cm]%
\gamma_{k}%
\end{array}
\right]  .
\end{align*}

\end{document}